\PassOptionsToPackage{table}{xcolor}
\documentclass[11pt,letterpaper]{style}
\usepackage[numbers]{natbib}
\usepackage{latexsym}
\usepackage{amsmath,amssymb,bm,mathtools}
\usepackage{graphicx,booktabs,array,tabularx,multirow,makecell,microtype}
\usepackage[table]{xcolor}
\usepackage{algorithm,algpseudocode}
\usepackage{enumitem}
\usepackage{placeins}
\hypersetup{pdftitle={SonicWorld: Source-Centric Latent Dynamics for Spatial Audio Reasoning},pdfauthor={Chenlong Wei, Hanling Wang, Mingjin Lou, Yufan Shi, Yandong Sun, Sheldon Yu, Xiaohui Zhu, Xiaobo Jin, Yiwei Wang}}
\newcommand{\method}{\textsc{SonicWorld}}

\newcommand{\R}{\mathbb{R}}

\definecolor{draftblue}{RGB}{44,86,119}
\newcommand{\pending}[1]{}

\newcolumntype{Y}{>{\raggedright\arraybackslash}X}
\usepackage{tcolorbox}
\tcbuselibrary{skins,breakable,listings}
\providecommand{\evaluationprompttextcolor}{black}
\newtcblisting{evaluationprompt}[1]{
  enhanced,
  listing only,
  title={#1},
  colback=SWBackground,
  colframe=SWBlue!45,
  colbacktitle=SWTableHighlight,
  coltitle=SWBlue,
  fonttitle=\bfseries\small,
  boxrule=0.65pt,
  arc=2mm,
  left=8pt,right=8pt,top=4pt,bottom=4pt,
  toptitle=3pt,bottomtitle=3pt,
  before skip=5pt,after skip=5pt,
  listing options={
    language={},numbers=none,frame=none,xleftmargin=0pt,xrightmargin=0pt,
    breakindent=0pt,
    breakautoindent=false,
    basicstyle=\ttfamily\fontsize{9}{11}\selectfont\color{\evaluationprompttextcolor},
    breaklines=true,
    breakatwhitespace=true,
    columns=fullflexible,
    keepspaces=true,
    showstringspaces=false,
    aboveskip=0pt,belowskip=0pt,
    literate={°}{{\textdegree}}1
  }
}

\usepackage{sonicworld_layout}
\title{SonicWorld: Source-Centric Latent Dynamics for Spatial Audio Reasoning}
\runningtitle{SonicWorld: Source-Centric Latent Dynamics for Spatial Audio Reasoning}
\author{%
  {\fontsize{11}{15}\selectfont\bfseries
  Chenlong Wei\textsuperscript{*1}\quad
  Hanling Wang\textsuperscript{*2}\quad
  Mingjin Lou\textsuperscript{3}\quad
  Yufan Shi\textsuperscript{2}\quad
  Yandong Sun\textsuperscript{4}\\
  Sheldon Yu\textsuperscript{2}\quad
  Xiaohui Zhu\textsuperscript{1}\quad
  Xiaobo Jin\textsuperscript{1}\quad
  Yiwei Wang\textsuperscript{\textdagger5}\par}
  \vspace{4pt}
  {\fontsize{10}{15}\selectfont\normalfont
  \textsuperscript{1}Xi'an Jiaotong-Liverpool University\\
  \textsuperscript{2}University of California, San Diego\quad
  \textsuperscript{3}Imperial College London\\
  \textsuperscript{4}Lyuliang University\quad
  \textsuperscript{5}Southeast University\par}
  {\centering\rmfamily * Equal contribution\quad\textdagger\ Corresponding author\par}
  {\normalfont\ttfamily\fontsize{8.5}{11}\selectfont
    \href{mailto:Chenlong.Wei23@student.xjtlu.edu.cn}{Chenlong.Wei23@student.xjtlu.edu.cn}\quad
    \href{mailto:haw168@ucsd.edu}{haw168@ucsd.edu}\\
   \href{mailto:wangyw.evan@gmail.com}{wangyw.evan@gmail.com}\\
  }
}
\newcommand{\SonicWorldProjectURL}{https://sonic-world.github.io/}
\begin{document}
% 先提出动态混合声场中的源归属和时间状态问题，
% 再介绍数据集与评测，再介绍维护声源状态的框架及其语言接口，最后阐明实验结果的提升
\begin{abstract}
Dynamic spatial audio reasoning requires a model to associate changing acoustic observations with the correct source and recover its spatial state at a queried time or interval, a challenge intensified by overlapping speech and source motion. 
%sonic bench
To study this problem, we introduce \textbf{Sonic-Bench}, comprising 2.8K first-order Ambisonics clips and 28.8K target-specific question--answer pairs across 480 simulated indoor scenes. 
The benchmark covers localization, distance estimation, and motion reasoning in single- and dual-source settings. 
% method 
We further propose \textbf{SonicWorld}, which integrates physical, learned acoustic, and semantic features into persistent source states through learned transitions and recurrent updates, and projects them into an Omni LLM for spatial reasoning. 
%result
Compared with Spatial-Omni on the Sonic-Bench test split, SonicWorld improves direction accuracy from 31.98\% to 34.69\% and reduces azimuth error from $57.22^\circ$ to $48.01^\circ$. In zero-shot transfer to Spatial LibriSpeech, SonicWorld retains strong localization performance, achieving 73.98\% direction accuracy and $19.29^\circ$ azimuth error, compared with 27.14\% and $70.74^\circ$ for Spatial-Omni. 
%code
% Code is available at: \url{https://anonymous.4open.science/r/2C38/}.
\end{abstract}

\maketitle
\suppressfloats[t]

\begin{figure*}[t]
    \centering
    \includegraphics[width=\linewidth]{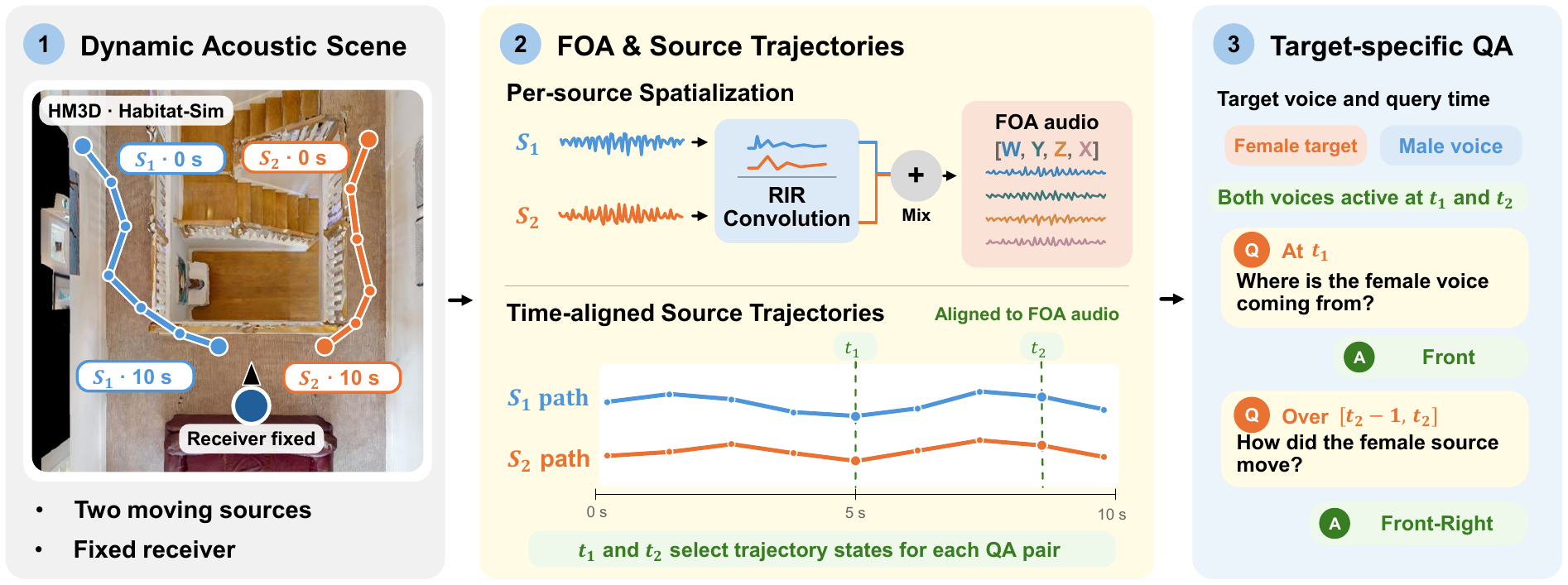}
    \caption{\textbf{Task overview of Sonic-Bench.}
    Given an FOA recording, the task is to infer a specified source's
    spatial attributes at a queried time or its movement direction
    over a queried interval. Time-aligned source trajectories provide
    ground-truth answers, illustrated here with two moving sources
    and a fixed receiver.}
    \label{fig:data_pipeline}
\end{figure*}
\section{Introduction}
\label{sec:introduction}
% Paragraph 1: Background
Recent Large Audio-Language Models (LALMs) and Omni LLMs have demonstrated growing capabilities in recognizing and reasoning about audio content \citep{tang2024salmonn,xu2025qwen25omni,guo2026survey}. However, understanding an auditory scene requires more than identifying what is heard: a model must also determine where each sound originates and how its location changes over time. Spatial audio preserves directional, distance, motion, and environmental cues beyond sound content, providing the perceptual basis for understanding dynamic 3D auditory scenes \citep{zhu2025asaudio}.

% Paragraph 2: Task gap
Spatial-audio benchmarks have expanded from localization to
multi-source relations and motion reasoning
\citep{starbench,spatialaqa}.
However, answering where a particular source is at a given
moment or how it moves over an interval calls for jointly
resolving source identity and temporal context.
We therefore study \emph{dynamic source-specific spatial
reasoning}: given a spatial-audio sequence and a query
identifying a source and a timestamp or interval, 
the model is asked to infer the source's location, distance, or movement
direction. 
Answering these queries requires linking acoustic evidence to the same source across time, yet most existing models treat spatial cues frame-wise or source-agnostically and thus struggle to maintain source identity under overlap and motion.

% method gap
Current approaches to spatial-audio reasoning typically
incorporate binaural features \citep{bat}, FOA intensity-vector
representations \citep{tangspatial,jaeger}, or parallel spatial
branches \citep{spatialomni} into language models.
These approaches make spatial cues accessible to language
models, yet still struggle to associate evolving observations with a specified source and recover its state at an arbitrary queried time or interval
\citep{sun2025spatialblindspot,spatialaqa}.
Inspired by structured trajectory modeling in audio-visual
systems \citep{savvy}, object-centric dynamics
\citep{wu2023slotformer}, and latent dynamics and world
modeling
\citep{hafner2019planet,wang2025avwm},
we organize acoustic observations into persistent source states
that capture spatial evolution for source- and time-specific
question answering.

% ours
% Benchmark: data and evaluation for source- and time-specific reasoning.
To support learning and evaluation of this task, we introduce
\textbf{Sonic-Bench}, comprising 2.8K FOA clips and
28.8K target-specific question--answer pairs across
480 simulated indoor scenes.
It covers stationary and moving speech in single- and
dual-source settings, with scene-disjoint training and test splits.
Each question specifies a target source and a timestamp or
interval for localization, distance estimation, or motion reasoning.
% Framework: how source states connect acoustic perception with language.
We further propose \textbf{SonicWorld}, a framework that maintains persistent source states using physical, learned acoustic, and semantic evidence.
Learned state transitions and recurrent observation updates model their temporal evolution, while trajectory supervision and masked-state prediction ground the states in source geometry.
A spatial projector converts the state sequences into tokens that the language model reads alongside audio and question tokens.
% 三项贡献：
The main contribution of this paper can be summarized as follows:
\begin{itemize}[leftmargin=2em]
    \item
    We formulate dynamic spatial audio reasoning as source-centric
    state inference and introduce \textbf{Sonic-Bench} for
    target- and time-specific localization, distance estimation,
    and motion reasoning.
    The benchmark pairs FOA recordings with full-clip source
    trajectories across single- and dual-source scenes, with
    scene-disjoint training and test splits.
    \item 
    We propose \textbf{SonicWorld}, a framework that learns persistent source states through predictive state transitions and recurrent observation updates, and projects the resulting state sequences into a language model for source- and time-specific spatial question answering.
    \item 
    Experiments demonstrate improved source localization on
    Sonic-Bench and consistent zero-shot direction and azimuth
    gains on Spatial LibriSpeech across three language backbones.
   
\end{itemize}

\begin{figure*}[t]
\centering\includegraphics[width=\linewidth]{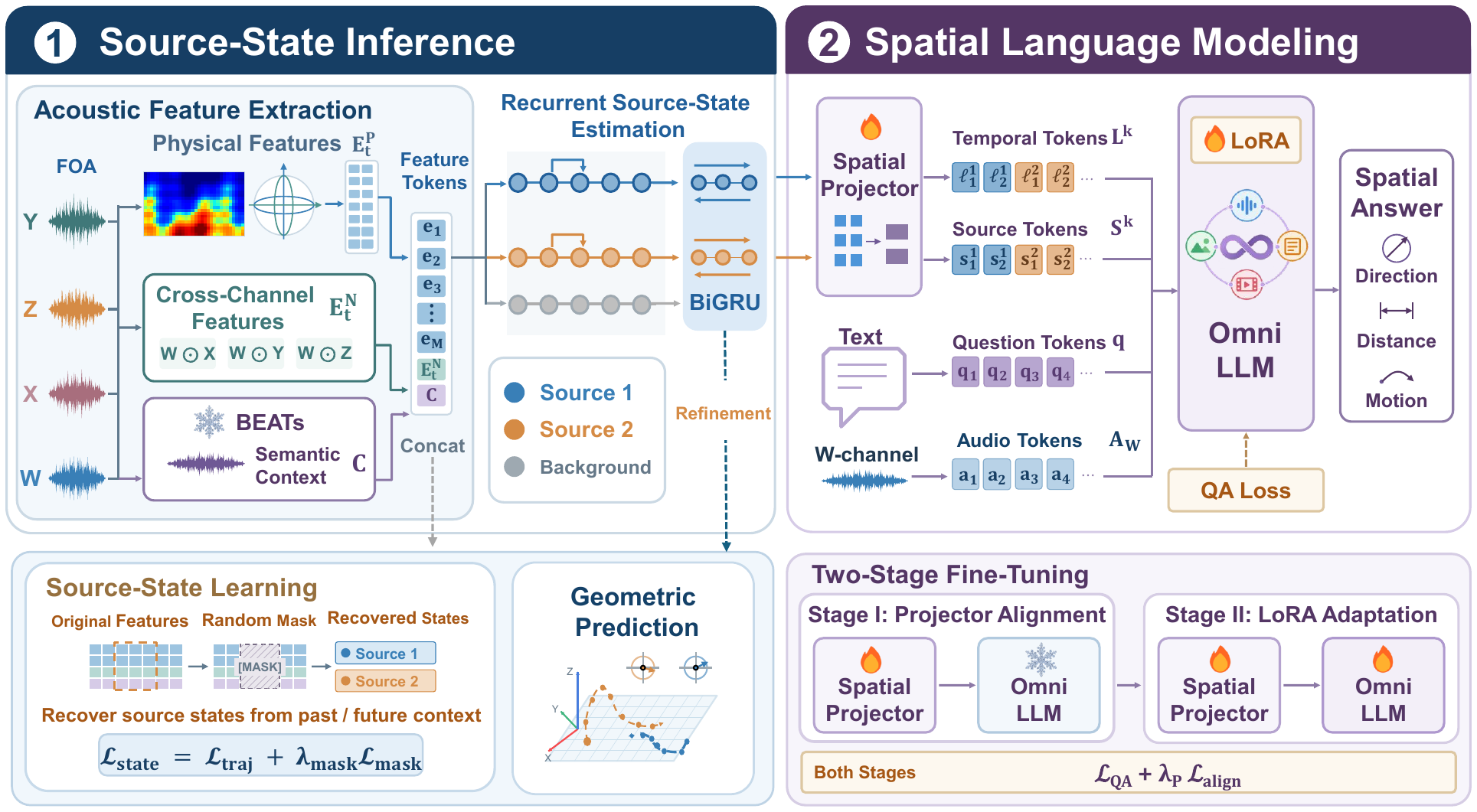}
\caption{\textbf{SonicWorld architecture.} Physical, neural, and frozen-BEATs semantic evidence are concatenated before recurrent source binding. Smoothed states and decoded geometry are projected into spatial tokens, which join audio and question tokens at the language model.}
\label{fig:architecture}
\end{figure*}
\section{Related Work}

\label{sec:related}
\subsection{Spatial Audio Perception and Dynamic Acoustic Scenes}
% spatial audio 的表示形式与基础空间感知任务
% 说明 binaural / microphone array / Ambisonics 如何携带空间信息
% 回顾 SELD、overlap 和 distance estimation，建立 spatial perception 的研究脉络
Spatial audio encodes scene geometry through binaural recordings, microphone arrays, and Ambisonics, using cues such as inter-channel timing, level differences, and sound-field structure \citep{zhu2025asaudio,grumiaux2022ssl_survey}.
% A central task is sound event localization and detection (SELD), which jointly recognizes active events and estimates their directions \citep{seldnet,shimada2021accdoa,adavanne2019tracking}.
% Subsequent work has extended SELD to handle overlapping same-class sources \citep{multiaccdoa} and to include source distance estimation \citep{dcase2024}.
A central task is sound event localization and detection (SELD),
which jointly recognizes active sound events and estimates
their directions of arrival \citep{seldnet,shimada2021accdoa}.
Research on SELD also addresses moving-source tracking
\citep{adavanne2019tracking}, localization of overlapping
same-class sources \citep{multiaccdoa}, and source distance
estimation \citep{dcase2024}.
For FOA signals, intensity vectors additionally provide physically grounded directional evidence from the multichannel sound field \citep{perotin2019crnn_doa,tangspatial,nguyen2022salsa}.
% 介绍真实空间音频数据与仿真平台
% 说明真实数据提供自然混响和干扰，而仿真提供精确 source geometry / trajectory supervision；
% 最后过渡到 dynamic / multi-source acoustic scenes。
Progress has also been supported by real and simulated spatial-audio environments \citep{evers2020locata,politis2021dynamic_interferers,chen2020soundspaces}.
STARSS23 provides multichannel recordings with spatiotemporal annotations under natural interference and source overlap \citep{starss23}.
SoundSpaces 2.0 and Habitat-Sim instead offer controllable acoustic rendering in scanned indoor scenes with known geometry and source locations \citep{soundspaces,habitatsim}.
While localization and tracking recover source geometry over time, they primarily operate at the frame or event level. We instead study source- and time-conditioned spatial reasoning, where a model must recover the spatial state of a specified source at a queried time or interval.
% % 转折嗷转折
% % While these resources advance acoustic perception, their frame- or
% % event-level tasks do not directly assess source-specific reasoning
% % that requires tracking source identities and evolving spatial states
% % across time.
% \textcolor{red}{While these resources advance spatial acoustic perception and
% tracking, they primarily formulate predictions at the frame or
% event level.
% We instead study source- and time-conditioned spatial reasoning,
% where the model is given a target source and a timestamp or
% interval and must recover the corresponding spatial state.}

%===================================================
\subsection{Audio, Omni, and Spatial Audio Language Models}
% 从LALM 到 Omni LLM。
% semantic audio understanding
Large Audio-Language Models connect pretrained audio encoders with
language models for audio understanding and reasoning \citep{chu2024qwen2_audio,gong2024ltu}.
SALMONN establishes this general architecture \citep{tang2024salmonn},
while Kimi-Audio and Audio Flamingo scale audio instruction following
\citep{kimi,kong2024audio_flamingo,af3}.
Omni models, including Qwen2.5-Omni and Qwen3-Omni, further integrate
audio, vision, and language within unified multimodal models
\citep{xu2025qwen25omni,qwen3}.
Together, these models provide strong semantic representations but do
not explicitly model spatial scene dynamics.
% Spatial Audio LLM 的主要技术路线及其共同局限。
% 但在动态多声源条件下的 source-specific temporal reasoning 仍然不足。
% Spatial-audio language models address this gap by incorporating explicit
% spatial cues.
Spatial audio–language alignment has also been explored via contrastive learning (e.g., ELSA \citep{devnani2024elsa} and SALM \citep{hu2025salm}). Spatial-audio LLMs further incorporate explicit spatial cues for question answering.
BAT learns binaural representations for localization and spatial
question answering \citep{bat}; FOA-based methods use intensity-vector
features \citep{tangspatial}; and JAEGER introduces learned Neural
Intensity Vectors for 3D audio-visual grounding \citep{jaeger}.
Spatial-Omni adds a parallel FOA branch for direction, distance, motion,
and multi-source reasoning \citep{spatialomni}.
% 转折
While these approaches expose spatial cues to language models, they do not maintain persistent source states across time. We organize successive acoustic observations into such states and project their temporal evolution into the language model for source- and time-specific spatial question answering.

% While these resources support increasingly complex acoustic perception, existing tasks primarily evaluate frame- or event-level predictions rather than tracking the identity and evolving state of individual sources for queries at specific times or intervals.
% They therefore provide a foundation for, but do not directly address, source-specific dynamic spatial reasoning.

% 1.
% Yet dynamic multi-source scenes remain challenging because 
% source interference and changing spatial cues require the model to
% maintain the evolving state of a specified source over time.   
% 2.

% \textcolor{red}{Spatial audio--language alignment has also been explored through
% contrastive learning.
% ELSA learns spatially aware audio--text embeddings
% \citep{devnani2024elsa}, while SALM separates semantic and spatial
% components in structured representations \citep{hu2025salm}.
% Spatial-audio LLMs incorporate explicit spatial cues for
% question answering.}

% 1.
% While localization and tracking estimate source geometry
% over time, our task couples these estimates with language
% queries specifying a target source and a timestamp or interval.
% The goal is to recover the target's spatial attributes at the
% queried time or its movement direction over the queried interval.
% 2.

% 1.
% Beyond incorporating spatial cues, we focus on organizing
% successive observations into persistent source states and
% exposing their temporal evolution to the language model for
% source- and time-specific spatial question answering.
% 2. 
% Dataset and benchmark revision, 2026-09-13.

\section{Sonic-Bench}

\subsection{Task Definition}
\label{sec:task_definition}
% task definition
Given a ten-second FOA recording $\mathbf{X}$ and a query $q$
identifying a source and a timestamp or interval, the task is to
infer its spatial state in a fixed receiver-centered frame, as shown in Figure~\ref{fig:data_pipeline}.
We consider five query types: \textit{position direction}
(eight sectors), \textit{azimuth}, \textit{elevation}, and
\textit{distance} at time $t$, and \textit{movement direction}
(eight directions or \textit{Stationary}) based on horizontal
net displacement over $[t-1,t]$.
%source
These queries span six source configurations: two single-source
settings with a stationary or moving target, and four dual-source
settings covering all target/interferer motion combinations.

% qa generation
Each clip has one designated target and two query times selected using
speech-activity annotations.
At each query time, we instantiate five task-specific questions
identifying the target voice and a timestamp or one-second interval,
with answers derived from the target trajectory, yielding ten QA
pairs per clip.
QA construction details and examples appear in
Appendix~\ref{app:dataset}, and evaluation metrics in
Appendix~\ref{app:results}.

\subsection{Data Collection}
\label{sec:data_collection}
Sonic-Bench evaluates source-specific spatial reasoning in dynamic single- and multi-source scenes. It contains 2,880 ten-second FOA clips, 4,800 time-aligned
source trajectories, and 28,800 QA pairs across 480 simulated
indoor scenes.
The scene-disjoint split comprises 2,400 training clips from
400 scenes and 480 test clips from 80 scenes.

% 交代语音、三维场景与仿真平台的来源，并定义接收器和声源的运动方式。
\paragraph{Scene and Speech Construction.}
We use speech from Mini LibriSpeech,
a subset of the LibriSpeech corpus \citep{librispeech},
and indoor scene meshes from HM3D \citep{hm3d}, configuring
source and receiver poses in Habitat-Sim \citep{habitatsim}.
The receiver has a fixed position and orientation; sources
either remain stationary or follow navigable trajectories.
Dual-source clips pair one female and one male voice, with one
designated as the target and the other as the interferer.

% 按 RIR 生成、语音空间化、多源混合、轨迹标注的顺序介绍声学合成
\paragraph{Acoustic rendering and annotations.}
We render source-specific FOA room impulse responses (RIRs) along
each trajectory using Habitat-Sim's acoustic propagation
engine \citep{soundspaces}, convolve them with dry speech,
and sum the spatialized signals into an FOA mixture.
For each source, we retain time-aligned positions, azimuth,
elevation, distance, and speech activity for QA generation.

% Dataset protocol precedes the method; revision 2026-09-12.

\section{SonicWorld}
\label{sec:method}

Given an FOA waveform $X\in \mathbb{R}^{4\times N}$ in channel order
$[W,Y,Z,X]$ and a question $q$, SonicWorld answers spatial
questions through persistent source-state inference
(Figure~\ref{fig:architecture}).
We combine physical spatial cues, learned waveform features,
and a clip-level semantic summary from frozen BEATs~\citep{beats}.
A recurrent estimator associates these features with individual
sources and updates their states over time.
During training, we randomly mask temporal feature spans and
supervise source-geometry recovery from surrounding context.
A projector converts the state sequences into local temporal
tokens and source summaries, which the language model reads
alongside native audio tokens and the question.
Architecture details are provided in Appendix~\ref{app:model}.

\subsection{Source-State Inference}
\label{sec:world}

\paragraph{Acoustic feature extraction.}
The physical branch combines channel log-magnitudes with active
and reactive intensity and inter-channel coherence from
short-time Fourier transforms of $X$.
Intensity uses the real and imaginary parts of the cross-spectra
between $W$ and each directional channel, normalized by
$W$-channel power; coherence measures channel-pair consistency.
A fixed Mel filterbank and temporal resampling yield
$\Phi\in\mathbb{R}^{T\times M\times13}$, with $T$ time steps
and $M$ Mel bands. A two-dimensional convolutional network
$f_{\mathrm{phys}}$ produces physical tokens:
\begin{equation}
E_t^P=\bigl[f_{\mathrm{phys}}(\Phi)\bigr]_t
\in\mathbb{R}^{M\times d_e},
\label{eq:physical_features}
\end{equation}
where $d_e$ is the token feature dimension.
These tokens carry normalized active-intensity directions and
coherence-derived reliability for geometric aggregation.

In parallel, a shared one-dimensional convolutional network
encodes each channel $c\in\{W,X,Y,Z\}$ into $H_c$.
Following Neural IV \citep{jaeger}, we concatenate elementwise
channel interactions along the feature dimension and map them
to one neural token per step:
\begin{align}
H_{\mathrm{int}}
&=\begin{aligned}[t]
\operatorname{Concat}(&H_W\odot H_X,H_W\odot H_Y,\\
&H_W\odot H_Z),
\end{aligned}
\label{eq:neural_interactions}\\
E_t^N
&=\bigl[g_{\mathrm{neural}}(H_{\mathrm{int}})\bigr]_t
\in\mathbb{R}^{1\times d_e}.
\label{eq:neural_features}
\end{align}
Here, $\odot$ denotes elementwise multiplication;
$g_{\mathrm{neural}}$ applies a two-layer MLP followed by
linear interpolation to $T$ steps.

Frozen BEATs \citep{beats} extracts $J$ patches of width $d_B$
from $W$, forming $F\in\mathbb{R}^{J\times d_B}$.
We normalize and project them to
$\widetilde F\in\mathbb{R}^{J\times d_e}$.
A learned query $q_c\in\mathbb{R}^{1\times d_e}$ pools these
patches through attention into a normalized semantic token $C$.
At each step, we concatenate the $M$ physical tokens, one neural
token, and the shared semantic token:
\begin{align}
C&=\operatorname{Attention}(q_c,\widetilde F,\widetilde F)
\in\mathbb{R}^{1\times d_e},
\label{eq:semantic_pooling}\\
E_t&=\operatorname{Concat}_{\mathrm{token}}(E_t^P,E_t^N,C)
\in\mathbb{R}^{(M+2)\times d_e}.
\label{eq:evidence_concat}
\end{align}

\paragraph{Recurrent source-state estimation.}
We associate $E_t$ with $K$ recurrent source slots and one
background slot, following sequential slot-based models
\citep{savi,elsayed2022savipp}.
Each slot $k$ maintains a $d_s$-dimensional state
$z_t^k\in\mathbb{R}^{d_s}$ at time $t$, initialized from
a learned $z_0^k$; the background slot is indexed by $K+1$.
Before reading $E_t$, a shared two-layer feed-forward network
$f_{\mathrm{trans}}$ predicts the prior:
\begin{equation}
\bar z_t^k=\operatorname{LayerNorm}
\left(z_{t-1}^k+f_{\mathrm{trans}}(z_{t-1}^k)\right).
\label{eq:prior}
\end{equation}
For token $E_{t,m}\in\mathbb{R}^{d_e}$, $m=1,\ldots,M+2$,
$W_Q\in\mathbb{R}^{d_s\times d_s}$ maps the prior to a query,
while $W_K,W_V\in\mathbb{R}^{d_s\times d_e}$ map tokens to
keys and values.
The scores form
$\boldsymbol{\ell}_t\in\mathbb{R}^{(K+1)\times(M+2)}$.
Following Slot Attention \citep{slotattention}, we apply softmax
across slots, then normalize each slot's weights over tokens:
\begin{align}
\ell_{t,k,m}
&=\frac{(W_Q\bar z_t^k)^\top(W_KE_{t,m})}{\sqrt{d_s}},
\label{eq:slot_compatibility}\\
A_{t,k,m}
&=\operatorname{softmax}_{k}(\ell_{t,k,m}),
\nonumber\\
\widetilde A_{t,k,m}
&=\frac{A_{t,k,m}}{\sum_{m'=1}^{M+2}A_{t,k,m'}}.
\label{eq:slot_assignment}
\end{align}
The resulting weights form a feature summary
$o_t^k\in\mathbb{R}^{d_s}$, which a gated recurrent unit
(GRU) \citep{cho2014gru} combines with the prior:
\begin{align}
o_t^k
&=\sum_{m=1}^{M+2}\widetilde A_{t,k,m}W_VE_{t,m},
\label{eq:slot_observation}\\
z_t^k
&=\operatorname{LayerNorm}
\left(\operatorname{GRU}(o_t^k,\bar z_t^k)\right).
\label{eq:slot_update}
\end{align}
Updated states pass to the next step, and the $K$ source
sequences are retained for refinement.

\paragraph{State refinement and geometric prediction.}
For each source, a bidirectional GRU aggregates earlier and later
states in $z_{1:T}$.
A linear projection $f_{\mathrm{ctx}}$ maps its output at $t$
to context $c_t\in\mathbb{R}^{d_s}$.
A gate conditioned on $z_t$ and $c_t$ controls the context
added to the recurrent state:
\begin{equation}
\begin{aligned}
c_t&=f_{\mathrm{ctx}}
\left(\bigl[\operatorname{BiGRU}(z_{1:T})\bigr]_t\right),
\\
g_t&=\sigma\left(f_{\mathrm{gate}}([z_t;c_t])\right),\\
h_t&=\operatorname{LayerNorm}(z_t+g_t\odot c_t)
\in\mathbb{R}^{d_s}.
\end{aligned}
\label{eq:temporal_smoothing}
\end{equation}
Here, $f_{\mathrm{gate}}$ is linear, $\sigma$ is sigmoid,
$[\cdot]_t$ selects time $t$, and $[\cdot;\cdot]$ concatenates
features.
Auxiliary heads predict unit direction $\hat u_t$, log-distance
$\hat d_t$, source presence, and positional uncertainty from $h_t$
for trajectory supervision. With $\hat r_t=\exp(\hat d_t)$, the receiver-centered position is
\begin{equation}
\hat p_t=\hat r_t\hat u_t\in\mathbb{R}^{3}.
\label{eq:geometric_readout}
\end{equation}
% % {\color{red}
% The decoded position $\hat p_t$ supplies the uncertainty position loss $\mathcal L_{\mathrm{uncertainty}}$ in Eq.~\ref{eq:app_uncertainty_loss}.
% % }

\paragraph{Source-state Learning.}
One Hungarian assignment per clip matches predicted source
sequences to ground-truth trajectories.
Matched annotations supervise direction, distance, and presence.
% We also use an uncertainty-weighted position loss and supervise
% direction and distance decoded from the transition prior
% $\bar z_t^k$.
 We also supervise $\hat p_t$ with an uncertainty
position loss and supervise direction and distance decoded
from the transition prior $\bar z_t^k$.

The trajectory and combined pretraining objectives are
\begin{align}
\mathcal L_{\mathrm{traj}}
&=\mathcal L_{\mathrm{dir}}+\mathcal L_{\mathrm{dist}}
+\mathcal L_{\mathrm{presence}}
\nonumber\\
&\quad+\lambda_u\mathcal L_{\mathrm{uncertainty}}
+\lambda_{\mathrm{prior}}\mathcal L_{\mathrm{prior}},
\label{eq:trajectory_loss}\\
\mathcal L_{\mathrm{state}}
&=\mathcal L_{\mathrm{traj}}
+\lambda_{\mathrm{mask}}\mathcal L_{\mathrm{mask}}.
\label{eq:state_loss}
\end{align}
Loss definitions and masking settings are provided in
Appendix~\ref{app:training}.

\subsection{Spatial Language Modeling}
\label{sec:projector}

\paragraph{Spatial projector.}
A trainable projector maps source-state features and additional spatial features to frame features.
It compresses consecutive frame features into local tokens,
while learned queries summarize each complete source sequence.
Both token types are projected to the language embedding space
and combined with native $W$-channel audio tokens and the
question $q$ for answer generation.
Projector details are provided in Appendix~\ref{app:model}.

\paragraph{Two-stage fine-tuning.}

As shown in Figure~\ref{fig:architecture},
Stage~I trains the projector and auxiliary alignment heads with the LLM
fixed; Stage~II
jointly trains the projector and low-rank adaptation (LoRA) parameters in its attention
layers.
The source-state model, native audio encoder, and original LLM
weights remain frozen. Both stages optimize

\begin{equation}
\mathcal L_{\mathrm{LLM}}
=\mathcal L_{\mathrm{QA}}+\lambda_P\mathcal L_{\mathrm{align}},
\label{eq:lm}
\end{equation}
where $\mathcal L_{\mathrm{QA}}$ is cross-entropy over answer
tokens. 
% The alignment loss trains only the projector and its alignment
% heads to match the geometric predictions of the frozen
% source-state model.
Algorithm~\ref{alg:sonicworld_training} details the training procedure.

\begin{table*}[!t]
\centering\footnotesize\color{black}
\setlength{\tabcolsep}{1.8pt}
\renewcommand{\arraystretch}{1.05}
\caption{\textbf{Experimental results on Sonic-Bench.}
Spatial baselines and SonicWorld use Qwen2.5-Omni-7B.
Dir./Mot. are label accuracies (\%); Az./El. are MAE in degrees
and Dist. is MAE in meters. Best values in each column are bold.}
\label{tab:main_results}
\label{tab:qa_single}
\label{tab:qa_dual}
\label{tab:internomni_split}
\label{tab:direct_audio_split}

\begin{tabularx}{\linewidth}{
@{}l*{10}{>{\centering\arraybackslash}X}@{}
}
\toprule
& \multicolumn{5}{c}{Single source}
& \multicolumn{5}{c}{Dual source}\\
\cmidrule(lr){2-6}\cmidrule(l){7-11}
Model
& Dir.$\uparrow$ & Az.$\downarrow$ & El.$\downarrow$
& Dist.$\downarrow$ & Mot.$\uparrow$
& Dir.$\uparrow$ & Az.$\downarrow$ & El.$\downarrow$
& Dist.$\downarrow$ & Mot.$\uparrow$\\
\midrule
\multicolumn{11}{c}{\textbf{Closed-source models}}\\
\midrule
Gemini 3 Flash
& 12.81 & 88.55 & 4.40 & 1.11 & 7.81
& 15.47 & 88.23 & 4.49 & 2.71 & 8.28\\
Gemini 3.1 Pro
& 9.06 & 90.87 & 15.78 & 1.86 & 12.19
& 14.53 & 90.32 & 28.88 & 2.03 & 14.38\\
\midrule
\multicolumn{11}{c}{\textbf{Open-source models}}\\
\midrule
Qwen2.5-Omni-7B
& 12.50 & 91.18 & 84.91 & 98.83 & 51.88
& 12.66 & 88.33 & 83.58 & 34.30 & 48.59\\
InternOmni
& 12.81 & 93.72 & 22.65 & 40.58 & 8.13
& 12.34 & 92.90 & 32.51 & 43.16 & 5.94\\
Phi-4-MM-5.6B
& 13.44 & 81.21 & 3.29 & 1.93 & 6.25
& 12.66 & 79.90 & 4.13 & 2.26 & 5.31\\
Audio Flamingo 3
& 14.06 & 85.62 & 9.60 & 2.41 & 7.81
& 12.34 & 88.80 & 17.32 & 3.12 & 5.16\\
Kimi-Audio-7B-Instruct
& 12.50 & 91.78 & 1.58 & 2.20 & 35.31
& 11.72 & 92.23 & 3.89 & 1.54 & 23.75\\
\midrule
\multicolumn{11}{c}{\textbf{Spatial baselines}}\\
\midrule
BAT
& 8.75 & 87.51 & 0.94 & 0.68 & 51.25
& 11.09 & 90.34 & 0.95 & 0.64 & 52.66\\
Classical-IV
& 29.38 & 47.92 & 1.14 & 0.58 & 51.56
& 20.94 & 66.06 & 1.08 & 0.62 & 52.66\\
Neural-IV
& 10.63 & 84.00 & 0.99 & 0.59 & 50.31
& 11.88 & 87.83 & 1.21 & 0.67 & 52.34\\
Spatial-Omni
& 43.75 & 46.87 & 1.57 & 1.36 & 49.06
& 26.09 & 62.40 & 1.12 & 3.08 & 50.94\\
\midrule
\rowcolor{SWTableHighlight}
\textbf{\method{}}
& \textbf{50.00} & \textbf{30.85} & \textbf{0.91}
& \textbf{0.56} & \textbf{52.50}
& \textbf{27.03} & \textbf{56.59} & \textbf{0.93}
& \textbf{0.60} & \textbf{53.91}\\
\bottomrule
\end{tabularx}

\vspace{10pt}
\begingroup
\normalsize
\makeatletter
\def\@captype{figure}
\makeatother
\centering
\includegraphics[width=1\linewidth]{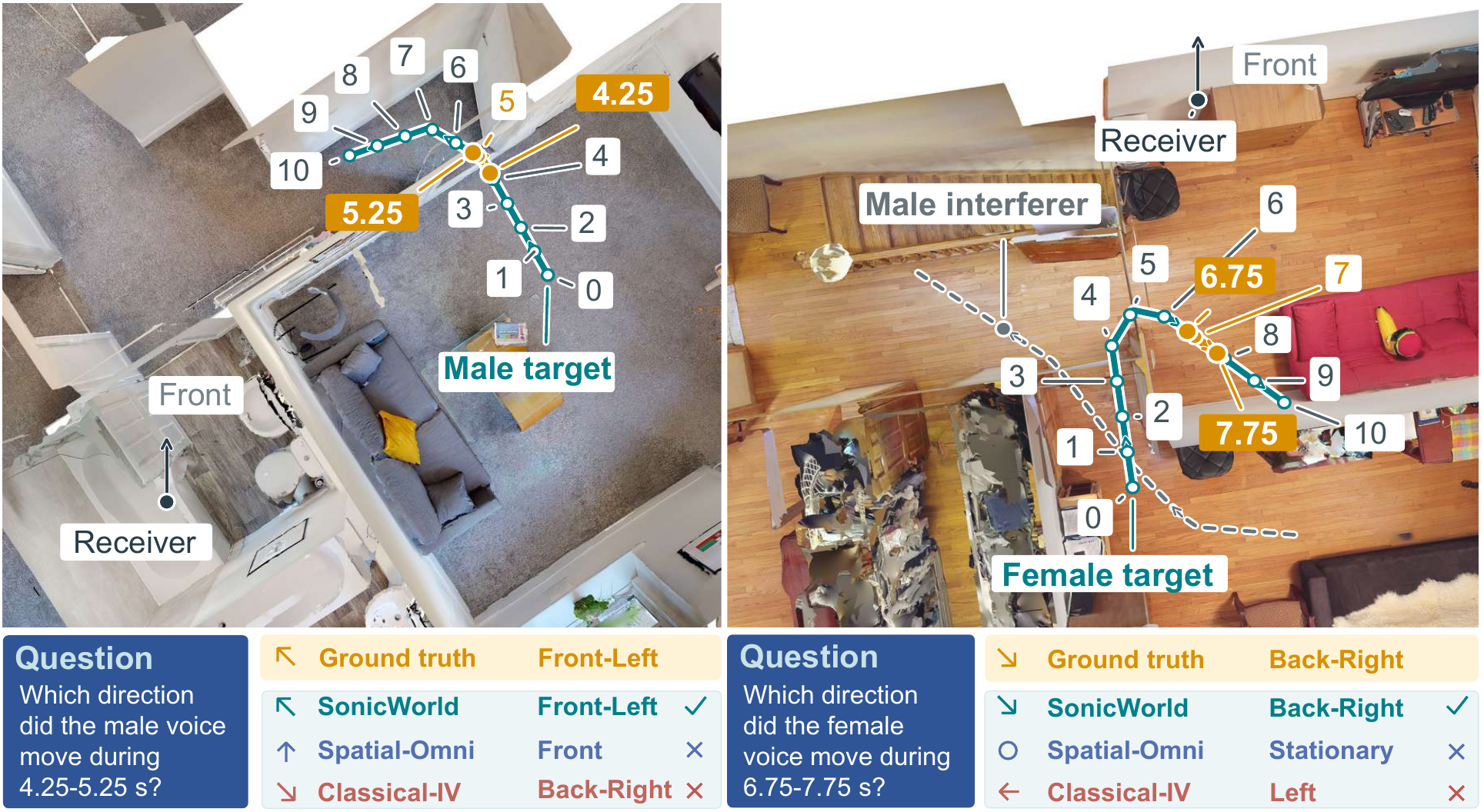}
\caption{Single-source (left) and dual-source (right) motion examples,
with queried intervals in orange. SonicWorld correctly predicts net
movement directions; Spatial-Omni and JAEGER (Classical-IV) do not.}
\label{fig:example}
\endgroup
\end{table*}
% \vspace{-15pt}
\section{Experimental Setup}
\label{sec:experiments}

\subsection{Training Details}
\label{sec:training}

We train SonicWorld on the training split of Sonic-Bench using Qwen2.5-Omni-7B
\citep{xu2025qwen25omni} as the backbone and pretrained BEATs \citep{beats}
for semantic audio feature extraction. 
We first train the source-state model for 25 epochs at a learning
rate of $10^{-4}$, then freeze it for the two fine-tuning stages.
The projector is trained for five epochs at a learning rate of
$10^{-4}$ with the LLM frozen. We then jointly optimize the projector
and LLM LoRA \citep{hu2021lora} parameters for three epochs, using the learning rates
of $3\times10^{-5}$ and $5\times10^{-5}$, respectively.
LoRA is applied to the LLM's attention layers with rank 16, a scaling parameter of 32, and dropout of 0.05.
Detailed optimization settings and loss weights are provided in
Appendix~\ref{app:training}.
% \vspace{-10pt}

% \begin{table}[!t]
% \centering\footnotesize\color{black}
% \setlength{\tabcolsep}{2pt}
% \renewcommand{\arraystretch}{1.10}
% \caption{\textbf{SonicWorld with different language backbones on Sonic-Bench.}
% Dir./Mot. are label accuracies (\%); Az./El. are MAE in degrees
% and Dist. is MAE in meters. Best values in each column are bold.}
% \label{tab:backbones}

% \begin{tabularx}{\linewidth}{
% @{}l*{10}{>{\centering\arraybackslash}X}@{}
% }
% \toprule
% & \multicolumn{5}{c}{Single source}
% & \multicolumn{5}{c}{Dual source}\\
% \cmidrule(lr){2-6}\cmidrule(l){7-11}
% Backbone
% & Dir.$\uparrow$ & Az.$\downarrow$ & El.$\downarrow$
% & Dist.$\downarrow$ & Mot.$\uparrow$
% & Dir.$\uparrow$ & Az.$\downarrow$ & El.$\downarrow$
% & Dist.$\downarrow$ & Mot.$\uparrow$\\
% \midrule
% Qwen2.5-Omni-3B
% & 50.00 & 31.98 & \textbf{0.90} & 0.55 & 52.19
% & 27.97 & 56.42 & \textbf{0.93} & \textbf{0.58} & 51.88\\
% Qwen2.5-Omni-7B
% & 50.94 & \textbf{30.70} & 0.91 & 0.57 & 55.00
% & 28.44 & \textbf{54.02} & 0.95 & 0.62 & \textbf{53.59}\\
% InternOmni
% & \textbf{51.88} & 36.49 & 0.92 & 0.61 & \textbf{55.94}
% & 30.16 & 62.49 & 0.93 & 0.62 & 52.19\\
% Phi-4-MM-5.6B
% & 48.75 & 32.03 & 0.92 & \textbf{0.53} & 53.44
% & \textbf{30.78} & 60.50 & 0.95 & 0.60 & 52.03\\
% \bottomrule
% \end{tabularx}
% \end{table}

\begin{table*}[htbp]
\centering\footnotesize\color{black}
\setlength{\tabcolsep}{3pt}
\renewcommand{\arraystretch}{1.05}
\caption{\textbf{Zero-shot transfer to SpatialSceneQA and SLS.}
Spatial baselines and SonicWorld use Qwen2.5-Omni-7B.}
\label{tab:zeroshot_results}

\begin{tabularx}{\linewidth}{
@{}l*{6}{>{\centering\arraybackslash}X}@{}
}
\toprule
& \multicolumn{2}{c}{SpatialSceneQA}
& \multicolumn{4}{c}{SLS}\\
\cmidrule(lr){2-3}\cmidrule(l){4-7}
Model
& Az.$\downarrow$ & El.$\downarrow$
& Dir.$\uparrow$ & Az.$\downarrow$
& El.$\downarrow$ & Dist.$\downarrow$\\
\midrule
\multicolumn{7}{c}{\textbf{Open-source models}}\\
\midrule
Qwen2.5-Omni-7B
& 60.47 & 80.14 & 12.69 & 90.26 & 88.47 & 55.04\\
InternOmni
& 80.54 & 78.05 & 12.38 & 89.69 & 57.59 & 98.83\\
Phi-4-MM-5.6B
& 61.86 & \textbf{18.31} & 12.61 & 89.43 & 7.08 & 1.99\\
Audio Flamingo 3
& 132.66 & 24.92 & 11.87 & 89.92 & 28.08 & 33.40\\
Kimi-Audio-7B-Instruct
& 55.37 & 35.44 & 11.63 & 90.74 & 9.41 & 3.90\\
\midrule
\multicolumn{7}{c}{\textbf{Spatial baselines}}\\
\midrule
BAT
& 125.37 & 20.51 & 11.85 & 90.75 & \textbf{7.07} & 0.84\\
Classical-IV
& 33.56 & 19.17 & 14.49 & 75.53 & 7.70 & 0.89\\
Neural-IV
& 88.33 & 18.49 & 12.71 & 89.38 & 7.74 & \textbf{0.78}\\
% SLS Spatial-Omni Dist.: retained previous-evaluation value 5.02 at the author's request.
% Current-evaluation raw MAE: 2.3769907297361543e40; other SLS cells use current results.
Spatial-Omni
& 108.28 & 22.84 & 27.14 & 70.74 & 16.64 & 5.02\\
\midrule
\rowcolor{SWTableHighlight}
\textbf{\method{}}
& \textbf{21.35} & 20.51 & \textbf{73.98} & \textbf{19.29} & 7.09 & 0.89\\
\bottomrule
\end{tabularx}
\end{table*} 

\subsection{Evaluation Metrics and Baselines}

\textbf{Evaluation metrics.}
\label{sec:evaluation}
We report classification accuracy (\%) for target-direction (Dir.)
and movement-direction (Mot.) questions on Sonic-Bench using exact
label matching. For azimuth (Az.), elevation (El.), and distance
(Dist.) estimation, we report mean absolute error (MAE), with angular
errors measured in degrees and distance errors in meters.
Azimuth error is computed as the shortest angular distance between
the predicted and ground-truth angles. We additionally report
accuracy at error thresholds of $20^\circ$, $1^\circ$, and 0.5\,m
for azimuth, elevation, and distance, respectively.
Detailed metric definitions and evaluation protocols are provided
in Appendix~\ref{app:results}.

\begingroup\emergencystretch=2em\relax
\textbf{Baselines.} We compare SonicWorld with the spatial audio baselines
Spatial-Omni \citep{spatialomni}, BAT \citep{bat},
JAEGER(Classical-IV) and JAEGER(Neural-IV) \citep{jaeger}. General-purpose multimodal baselines
include the open-weight models Qwen2.5-Omni-7B \citep{xu2025qwen25omni},
InternOmni \citep{chen2024far,chen2024internvl}, Audio Flamingo 3 \citep{af3},
Phi-4-MM-5.6B \citep{phi4mm} and Kimi-Audio-7B-Instruct \citep{kimi}, as well as the
closed-source models Gemini 3 Flash and Gemini 3.1 Pro \citep{team2023gemini}.
\par\endgroup

% \textbf{Zero-Shot Transfer Evaluation.}
% \label{sec:external_evaluation}
% We evaluate zero-shot transfer to selected test subsets of SpatialSceneQA \citep{jaeger} and Spatial LibriSpeech (SLS) \citep{spatial_librispeech2023}.
% We retain recordings of at least ten seconds and use their first ten seconds.
% On SpatialSceneQA, we evaluate azimuth and elevation separately, while SLS evaluates direction, azimuth, elevation, and distance for single-source speech.
% Table~\ref{tab:external_results} reports zero-shot localization results; sample selection, prompt construction, and source-count breakdowns are detailed in Appendix~\ref{app:results}.

% \setlength{\intextsep}{6pt}
% \setlength{\textfloatsep}{6pt}
% % Keep results-page spacing natural around the large tables.
% \raggedbottom
\section{Results}
\label{sec:results}

\subsection{Main results}

With the same Qwen2.5-Omni-7B backbone, SonicWorld improves
direction accuracy over Spatial-Omni from 43.75\% to 50.00\%
for single sources and from 26.09\% to 27.03\% for dual sources.
Azimuth MAE decreases from 46.87$^\circ$ to 30.85$^\circ$
and from 62.40$^\circ$ to 56.59$^\circ$, respectively.
These gains persist under source overlap, although their
magnitude is smaller. Dual-source distance error is
0.60\,m, compared with 0.62\,m for Classical-IV.

SonicWorld obtains 52.50\%/53.91\% movement-direction
accuracy on single/dual sources, versus BAT at 51.25\%/52.66\%.
Figure~\ref{fig:example} illustrates two cases where SonicWorld
correctly predicts the target's interval-specific movement,
while Spatial-Omni and JAEGER (Classical-IV) do not.
% backbone
Comparisons across language backbones are provided in
Appendix~\ref{app:backbones}.

\subsection{Zero-Shot Transfer}
\label{sec:zero_shot_transfer}
Zero-shot evaluation tests whether SonicWorld checkpoints
trained on Sonic-Bench transfer to static-source localization
without additional training.
The evaluation uses selected test subsets of
SpatialSceneQA \citep{jaeger} and Spatial LibriSpeech (SLS)
\citep{spatial_librispeech2023}, retaining recordings of at
least ten seconds and using their first ten seconds.
SpatialSceneQA tests azimuth and elevation separately,
while single-source SLS covers direction, azimuth,
elevation, and distance.
Appendix~\ref{app:results} details sample selection and
source-count breakdowns.
% ; Appendix~\ref{app:prompts} provides
% the task instructions and question prompts.

Table~\ref{tab:zeroshot_results} reports
zero-shot transfer results on two external datasets.
On SpatialSceneQA, SonicWorld reduces azimuth MAE by
80.3\% relative to Spatial-Omni, with improvements in
both single- and dual-source settings
(Appendix~\ref{app:results}).
On SLS, SonicWorld achieves the highest
direction accuracy (73.98\%) and lowest azimuth MAE
($19.29^\circ$) among the models reported in
Table~\ref{tab:zeroshot_results}.
\textcolor{black}{Zero-shot comparisons across the three
evaluated backbones are consolidated in
Table~\ref{tab:external_backbones}.}

\subsection{Ablation studies}

\newcommand{\ablOn}{{\fontencoding{U}\fontfamily{pzd}\selectfont\char52}}
\newcommand{\ablOff}{{\fontencoding{U}\fontfamily{pzd}\selectfont\char56}}
\begin{table*}[htbp]
\centering\footnotesize
\setlength{\tabcolsep}{3pt}\renewcommand{\arraystretch}{1.14}
\caption{\textbf{Module ablations with Qwen2.5-Omni-7B.}
\ablOn{} denotes enabled and \ablOff{} disabled.
}
\label{tab:ablations1}
\begin{tabular*}{\linewidth}{@{\extracolsep{\fill}}ccccrrrrrr@{}}
\toprule
Physical & Neural & BEATs & BiGRU & Dir.$\uparrow$ & Mot.$\uparrow$ & Az20$\uparrow$ & Az.$\downarrow$ & El.$\downarrow$ & Dist.$\downarrow$\\
\midrule
\ablOn & \ablOff & \ablOff & \ablOn & 28.33 & 52.40 & 25.73 & 55.68 & 0.97 & 0.62\\
\ablOn & \ablOn & \ablOff & \ablOn & 28.65 & 52.19 & 26.88 & 55.59 & 0.99 & 0.62\\
\ablOn & \ablOn & \ablOn & \ablOff & 32.19 & 51.98 & 33.13 & 48.11 & 0.96 & 0.63\\
\midrule
\ablOn & \ablOn & \ablOn & \ablOn & \textbf{34.69} & \textbf{53.44} & \textbf{34.90} & \textbf{48.01} & \textbf{0.92} & \textbf{0.59}\\
\bottomrule
\end{tabular*}
\end{table*}

\begin{table*}[htbp]
\centering
\caption{\textbf{Ablations of source-state modeling with Qwen2.5-Omni-7B.}}
\label{tab:ablations2}
\footnotesize
\setlength{\tabcolsep}{4pt}
\begin{tabular*}{\linewidth}{@{\extracolsep{\fill}}lrrrrrr@{}}
\toprule
Variant & Dir.$\uparrow$ & Mot.$\uparrow$ & Az20$\uparrow$
 & Az.$\downarrow$ & El.$\downarrow$ & Dist.$\downarrow$ \\
\midrule

w/o transition MLP     & 33.44 & 51.46 & 32.71 & 49.94 & 0.96 & 0.64 \\     
w/o prior-state supervision & 24.48 & 51.67 & 22.19 & 61.25 & 1.05 & 0.65 \\
w/o masked-state loss    & 33.02 & 52.81 & 30.21 & 50.66 & 0.98 & 0.64 \\   
w/o persistent source slots & 33.65 & 52.81 & 25.94 & 54.13 & 1.09 & 0.68 \\
\midrule
Full model                   & \textbf{34.69} & \textbf{53.44} & \textbf{34.90} & \textbf{48.01} & \textbf{0.92}& \textbf{0.59} \\
\bottomrule
\end{tabular*}
\end{table*}

\paragraph{Effectiveness of Modules.}
The full model achieves the best results across all six
metrics in Table~\ref{tab:ablations1}. Physical-only and no-BEATs have
azimuth MAEs $7.67^\circ$ and $7.58^\circ$ higher than Full,
and Az20 scores 9.17 and 8.02 points lower, respectively.
No-BiGRU is close in azimuth MAE ($48.11^\circ$ vs. $48.01^\circ$),
but has lower Dir. (32.19\% vs. 34.69\%) and higher distance MAE
(0.63\,m vs. 0.59\,m).

\paragraph{Effectiveness of Source-state Modeling.}
As shown in Table~\ref{tab:ablations2}, \textcolor{black}{
removing prior-state supervision increases azimuth MAE by
$13.24^\circ$, highlighting the importance of geometrically
grounded priors.
Removing persistent source slots yields the highest elevation
and distance errors, while removing the transition MLP causes
the largest motion-accuracy drop (1.98 percentage points).
Without masked-state loss, Az20 decreases by 4.69 percentage
points, demonstrating the benefit of learning to recover
source geometry from temporal context.}

\FloatBarrier

\section{Conclusion}
% background
Dynamic spatial audio reasoning requires associating acoustic
observations with individual sources and tracking their spatial
evolution over time.
% bench+our solution
We introduce \textbf{Sonic-Bench}, which pairs FOA recordings
with time-aligned trajectories and source-specific spatial
questions in dynamic single- and multi-source scenes.
To address these tasks, we propose \textbf{SonicWorld}, which
integrates physical, learned acoustic, and semantic features
into persistent source states through trajectory supervision
and masked-state prediction, and connects these states to
an Omni LLM for spatial question answering.
%result
Experiments show that persistent source-state modeling improves
source-specific localization and motion reasoning on Sonic-Bench,
while zero-shot evaluations demonstrate transferable localization
gains across different language backbones.
These results suggest that maintaining explicit source-level
temporal states provides an effective interface between spatial
audio perception and language-based reasoning, enabling models to
reason about evolving auditory scenes beyond individual acoustic
observations.

% \section*{Limitations}
% Our current study focuses on simulated indoor environments.
% While this setting enables controlled evaluation of 
% spatial reasoning, broader acoustic conditions, source categories,
% and denser multi-source scenes remain to be explored.
% The receiver is stationary, leaving
% receiver motion and changing egocentric reference frames outside
% the current scope.
% In addition, the bidirectional temporal module operates on the
% complete audio clip, making the present implementation better
% suited to offline analysis than streaming perception.
% Future work could explore real-world recordings, moving receivers,
% and causal state updates, and extend persistent source-state
% modeling toward action-conditioned prediction of acoustic
% observations for robotic active listening.
\section*{Limitations}
This work has several limitations:
\begin{itemize}
    \item \textbf{Simulated indoor environments.}
    Our study focuses on simulated indoor scenes.
    Broader acoustic conditions, source categories, and denser
    multi-source scenes remain to be explored, along with
    validation on real-world recordings.

    \item \textbf{Stationary receiver.}
    The receiver remains fixed, leaving receiver motion and
    changing egocentric reference frames outside the current scope.
    Future work could incorporate moving receivers and extend
    source-state modeling toward action-conditioned prediction
    of acoustic observations for robotic active listening.

    \item \textbf{Offline inference.}
    The bidirectional temporal module processes the complete
    audio clip, making the current implementation better suited
    to offline analysis than streaming perception.
    Causal state updates could support continuous auditory
    interaction.
\end{itemize}
% \section*{Ethics Statement}
% % [ETH-P1] 只写实际资源来源与任务相关的影响；不声称未经确认的伦理批准或新录音同意。
% The spatial recordings are synthesized from existing speech resources and scanned environments. Source attributes used in questions are inherited dataset labels. Spatial localization and trajectory inference may support assistive listening and embodied interfaces, while also creating risks of unconsented tracking and disclosure of conversational context. Releases must respect the licenses and access conditions of the speech, scenes, and pretrained models. Resource provenance is summarized in
% Appendix~\ref{app:dataset}.

% \section*{Reproducibility Statement}

% Dataset construction, spatial annotations, and task definitions
% are detailed in Appendix~\ref{app:dataset}.
% Appendix~\ref{app:model} describes the model architecture,
% while Appendix~\ref{app:training} specifies SonicWorld's
% training objectives and optimization settings.
% Evaluation protocols, metrics, and supplementary results
% are provided in Appendix~\ref{app:results}.
% Task instructions and question templates are presented in
% Appendix~\ref{app:prompts}.
% Code is available at
% \url{https://anonymous.4open.science/r/2C38/}.
% Sonic-Bench and model checkpoints are available at
% \url{https://doi.org/10.5281/zenodo.22814784}.

% \section*{AI Use Statement}

% Generative AI assistance was used for manuscript writing and language polishing, as well as for literature retrieval and discovery, including identifying potentially relevant related work.
% All AI-assisted text and suggested references were reviewed
% by the authors, and cited works were checked against their
% original sources.

\label{page:mainend}
\FloatBarrier
% Numeric bibliography style used by the reference template.
\bibliographystyle{unsrtnat}
\begingroup
% Allow long reference URLs to wrap without stretching surrounding text.
\expandafter\def\expandafter\UrlBreaks\expandafter{\UrlBreaks
  \do\a\do\b\do\c\do\d\do\e\do\f\do\g\do\h\do\i\do\j\do\k\do\l\do\m\do\n\do\o\do\p\do\q\do\r\do\s\do\t\do\u\do\v\do\w\do\x\do\y\do\z
  \do\A\do\B\do\C\do\D\do\E\do\F\do\G\do\H\do\I\do\J\do\K\do\L\do\M\do\N\do\O\do\P\do\Q\do\R\do\S\do\T\do\U\do\V\do\W\do\X\do\Y\do\Z
  \do\0\do\1\do\2\do\3\do\4\do\5\do\6\do\7\do\8\do\9}
\Urlmuskip=0mu plus 1mu\relax
\def\UrlBreakPenalty{100}
\bibliography{references}
\endgroup
% Let the appendix follow the references without an artificial blank remainder.
% \appendix
% \renewcommand{\thetable}{\thesection\arabic{table}}
% \renewcommand{\thefigure}{\thesection\arabic{figure}}
% \renewcommand{\theHtable}{appendix.\arabic{table}}
% \renewcommand{\theHfigure}{appendix.\arabic{figure}}
% \input{appendix/a_related_work}
% \input{appendix/b_model}
% \input{appendix/c_dataset}
% \input{appendix/d_training}
% \input{appendix/e_benchmark}
% \input{appendix/f_baselines}
% \input{appendix/g_results}
% \input{appendix/h_prompts}
% \input{appendix/i_licenses}

% 我先看你正文
\appendix

\numberwithin{equation}{section}
\counterwithin{table}{section}
\counterwithin{figure}{section}

\renewcommand{\thetable}{\thesection.\arabic{table}}
\renewcommand{\thefigure}{\thesection.\arabic{figure}}
\renewcommand{\theHtable}{appendix.\thesection.\arabic{table}}
\renewcommand{\theHfigure}{appendix.\thesection.\arabic{figure}}
\renewcommand{\theHequation}{appendix.\thesection.\arabic{equation}}

\clearpage
\section{Sonic-Bench}
\label{app:dataset}

% 语料来源、场景划分与语音复用在同一处说明。
\subsection{Data Sources and Scene-Level Split}
\label{app:speech_sources}
\label{app:dataset_splits}
Sonic-Bench combines speech from Mini LibriSpeech
\citep{librispeech}, distributed under CC BY 4.0,
with scanned indoor scene meshes from HM3D \citep{hm3d},
subject to the provider's access and licensing terms.
We partition the benchmark by scene: 400 scenes from the
HM3D training set form our training split, and 80 scenes
from its validation set form our test split.
Table~\ref{tab:scale} summarizes the resulting dataset.

\begin{table}[!htbp]
\centering\footnotesize\color{black}
\setlength{\tabcolsep}{2pt}\renewcommand{\arraystretch}{1.0}
\caption{Sonic-Bench statistics by split.
Each clip contains two target query times and five questions per query time.}
\label{tab:scale}\label{tab:sonic_split_summary}
\begin{tabularx}{\linewidth}{@{}Xccc@{}}
\toprule
\textbf{Unit} & \textbf{Train} & \textbf{Test} & \textbf{Total}\\
\midrule
Scenes & 400 & 80 & 480\\
Single-source clips & 800 & 160 & 960\\
Dual-source clips & 1,600 & 320 & 1,920\\
% FOA clips & 2,400 & 480 & 2,880\\
Source trajectories & 4,000 & 800 & 4,800\\
% Target query times & 4,800 & 960 & 5,760\\
% Annotated source states & 484,000 & 96,800 & 580,800\\
Question--answer pairs & 24,000 & 4,800 & 28,800\\
Audio duration & 6 h 40 min & 1 h 20 min & 8 h\\
\bottomrule
\end{tabularx}
\end{table}

We use one utterance per speaker from 24 speakers
(14 female-labeled, 10 male-labeled) in Mini LibriSpeech's
\texttt{train-clean-5} and \texttt{dev-clean-2} archives.
Utterances of at least 10\,s are converted to 16\,kHz mono
and randomly cropped to 10\,s, with approximately 10\,ms
fades at both ends.
Dual-source clips pair independently sampled female and male
speech, with target gender balanced in both single- and
dual-source settings.

\subsection{Comparison with Existing Resources}
Table~\ref{tab:dataset_comparison} summarizes representative
spatial-audio resources.
Sonic-Bench combines full-clip source trajectories with
target- and time-specific questions, covering stationary
and moving targets in single- and dual-source scenes
with all target/interferer motion combinations.

\begin{table*}[t]
\centering
\footnotesize
\setlength{\tabcolsep}{2.5pt}
\renewcommand{\arraystretch}{1.3}

\caption{\textbf{Comparison with spatial-audio resources.}
Scale uses each resource's reporting unit;
Spatial supervision refers to geometry available during
construction or training.}
\label{tab:dataset_comparison}

\begin{tabularx}{\linewidth}{
@{}
>{\raggedright\arraybackslash}p{.19\linewidth}
>{\raggedright\arraybackslash}p{.125\linewidth}
>{\centering\arraybackslash}p{.12\linewidth}
>{\centering\arraybackslash}p{.08\linewidth}
>{\raggedright\arraybackslash}p{.22\linewidth}
>{\raggedright\arraybackslash}X
@{}
}
\toprule
\textbf{Resource} &
\textbf{Format} &
\textbf{Scale} &
\textbf{Motion} &
\textbf{Spatial supervision} &
\textbf{Queries} \\
\midrule

SpatialSoundQA \citep{bat}
& Binaural
& 872K QA
& Static
& Static source geometry
& Source attributes / relations \\

SpatialSceneQA \citep{jaeger}
& FOA, RGB-D
& 61K samples
& Static
& Static 3D poses
& Source identity / visual grounding \\

SAVVY-Bench \citep{savvy}
& 7-ch audio, video
& 1,522 QA
& Static / Moving
& Camera-derived speaker tracks
& Sound event-anchored \\

STAR-Bench \citep{starbench}
& Binaural
& 2,353 QA
& Static / Moving
& Motion / trajectory QA labels
& Motion-pattern reasoning\\

SO-Bench \citep{spatialomni}
& FOA
& 7K QA
& Static / Moving
& Linked SO-Dataset tracks
& Task-dependent source / time \\

\midrule

\textbf{Sonic-Bench}
\textbf{(ours)}
& FOA
& 28.8K QA
& Static / Moving
& Full-clip source tracks
& Target- and time-specific \\

\bottomrule
\end{tabularx}
\end{table*}

% \paragraph{Split composition.}
% The training and test splits share all 24 speakers and original
% utterances. Of the 800 test source instances, 294 (36.75\%) use
% a dry-waveform crop that also appears in training.
% Scene geometries are disjoint, so this split evaluates transfer
% to unseen environments with shared speech material.

% 几何采样、运动配置与位置扇区配额集中说明。

\subsection{Scene Layouts and Source Trajectories}
The receiver remains at a fixed position and orientation.
Moving trajectories follow navigable shortest paths between sampled
endpoints and are resampled uniformly in cumulative path length.
This produces constant path speed, with direction changes at turns.
Stationary sources retain a single position throughout the clip.
The geometric constraints and rendering parameters are listed in
Table~\ref{tab:generation_settings}.

% \begin{table}[!htbp]
% \centering\footnotesize\color{black}
% \setlength{\tabcolsep}{2pt}\renewcommand{\arraystretch}{1.0}
% \caption{\textbf{Scene and acoustic generation parameters.}}
% \label{tab:generation_settings}
% \begin{tabularx}{\linewidth}{@{}>{\raggedright\arraybackslash}p{0.47\linewidth}X@{}}
% \toprule
% \textbf{Parameter} & \textbf{Setting}\\
% \midrule
% Clip duration; sample rate & 10 s; 16 kHz\\
% Geometric sampling & 12 Hz; 121 frames including endpoints\\
% Receiver / source height above NavMesh & 1.50 m / 1.55 m\\
% Source--receiver distance along the path & 1.5--4 m\\
% Moving path length & 2--5 m\\
% Minimum separation between sources & 0.8 m\\
% % Target-bearing constraint & One assigned $45^\circ$ sector throughout the path\\
% RIR candidate rate; displacement trigger & At most 4 Hz; 0.35 m\\
% RIR support; terminal fade & 0.25 s; 0.025 s\\
% Propagation components & Direct, indirect, diffraction, transmission\\
% Acoustic materials & Default uniform materials\\

% FOA output & Four-channel PCM16; 16 kHz; N3D metadata\\
% \bottomrule
% \end{tabularx}
% \end{table}
\begin{table}[!htbp]
\centering
\footnotesize
\color{black}
\setlength{\tabcolsep}{4pt}
\renewcommand{\arraystretch}{1.12}

\caption{\textbf{Scene and acoustic generation parameters.}}
\label{tab:generation_settings}

\begin{tabularx}{\linewidth}{
@{}
>{\raggedright\arraybackslash}p{0.40\linewidth}
>{\raggedright\arraybackslash}X
@{}
}
\toprule
\textbf{Parameter} & \textbf{Setting}\\
\midrule

Duration / sample rate
& 10\,s / 16\,kHz\\

Geometric sampling
& 12\,Hz; 121 frames\newline
  (including endpoints)\\

Receiver height
& 1.50\,m above NavMesh\\

Source height
& 1.55\,m above NavMesh\\

Source--receiver range
& 1.5--4\,m along the path\\

Moving-path length
& 2--5\,m\\

Source separation
& At least 0.8\,m\\

\midrule

RIR candidate rate
& At most 4\,Hz\\

Displacement trigger
& 0.35\,m\\

RIR support
& 0.25\,s\\

Terminal fade
& 0.025\,s\\

Propagation components
& Direct, indirect,\newline
  diffraction, transmission\\

Acoustic materials
& Default uniform materials\\

\midrule

FOA output
& Four-channel PCM16\\

FOA metadata
& N3D\\

\bottomrule
\end{tabularx}
\end{table}

The six target/interferer motion configurations in
Table~\ref{tab:conditions} are crossed with eight bearing sectors:
Front, Front-Left, Left, Back-Left, Back, Back-Right, Right, and
Front-Right. The receiver heading places the complete target
path within its assigned sector. Both position-direction questions
in a clip therefore have the same sector label; azimuth, elevation,
distance, and interval displacement are computed at their respective
query times. Figure~\ref{fig:bearing_sectors} visualizes the sector boundaries and the balanced clip distribution.

\begin{table*}[!htbp]
\centering
\begin{minipage}[c]{0.49\linewidth}
\centering
\includegraphics[width=\linewidth]{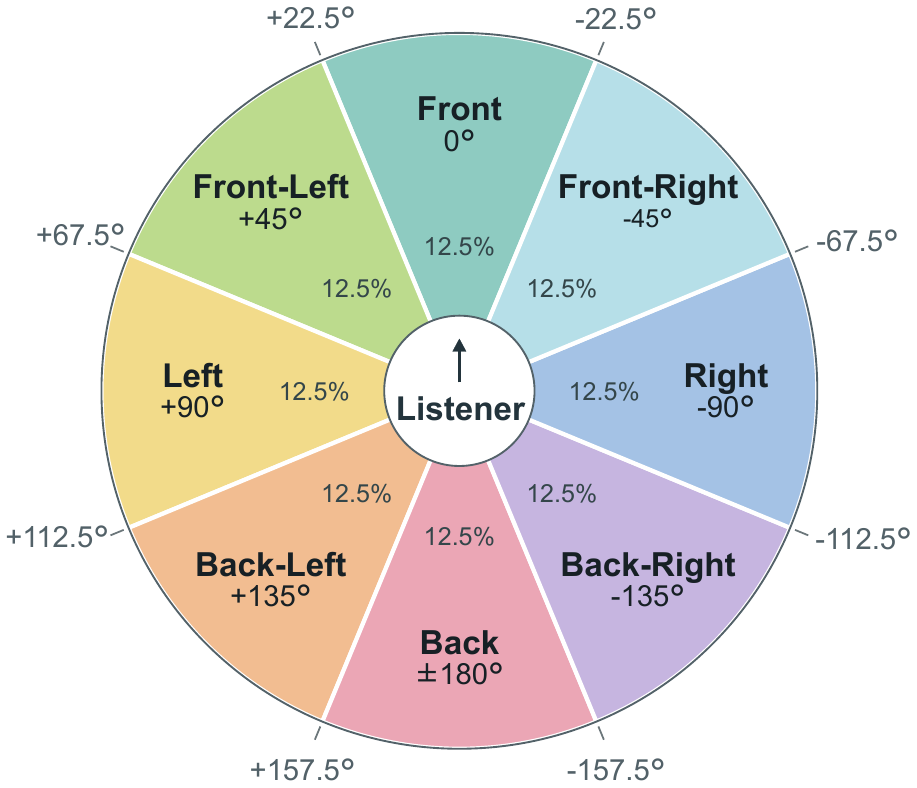}
\end{minipage}\hfill
\begin{minipage}[c]{0.49\linewidth}
\centering\footnotesize\color{black}
\setlength{\tabcolsep}{2pt}
\renewcommand{\arraystretch}{1.0}
\makeatletter
\def\@captype{table}
\makeatother
\caption{\textbf{Source-motion configurations.} Each row is balanced across eight target-bearing sectors.}
\label{tab:conditions}
\begin{tabularx}{\linewidth}{@{}Xllcc@{}}
\toprule
\textbf{Sources} & \textbf{Target} & \textbf{Interferer} & \textbf{Train} & \textbf{Test}\\
\midrule
Single & Moving & -- & 400 & 80\\
Single & Stationary & -- & 400 & 80\\
Dual & Moving & Moving & 400 & 80\\
Dual & Moving & Stationary & 400 & 80\\
Dual & Stationary & Moving & 400 & 80\\
Dual & Stationary & Stationary & 400 & 80\\
\bottomrule
\end{tabularx}
\end{minipage}
\makeatletter
\def\@captype{figure}
\makeatother
\caption{\textbf{Target-bearing sectors and clip distribution.} Each sector spans $45^\circ$ and contains 12.5\% of the clips in each split.}
\label{fig:bearing_sectors}
\end{table*}

% 按 RIR 更新、分段卷积和混合的实际顺序说明声学合成。
\subsection{Acoustic Synthesis}
\label{app:acoustics}
Habitat-Sim's geometry-based propagation engine renders a
source-specific FOA room impulse response at selected path positions
\citep{habitatsim,soundspaces}. At each candidate time, a new response
is computed when displacement from the previous acoustic keyframe
reaches 0.35\,m. Stationary sources generally use one response.
Each RIR is truncated to 0.25\,s and tapered over its final 0.025\,s.

Dry speech is partitioned at acoustic keyframes. Each segment is
convolved with its corresponding RIR; response tails are added where
they overlap. With dry waveform $x_s$, keyframe interval
$\mathcal I_{s,j}$, and channel-specific response $h_{s,c}^{(j)}$,
the mixture is
\begin{equation}
y_c[n]=G\sum_s\sum_j
\left(h_{s,c}^{(j)} * (x_s\mathbf 1_{\mathcal I_{s,j}})\right)[n].
\label{eq:dynamic_convolution}
\end{equation}
Responses are rotated and reordered to the receiver convention.
The shared gain $G$ limits the peak amplitude of all four channels
together, and the output is cropped to 160,000 samples.

% 将原模型附录和数据附录中的坐标定义合并。
\subsection{Coordinates and Source Annotations}
\label{app:coordinates}
All model positions are receiver-centered, with $+x$ forward,
$+y$ left, and $+z$ upward. The receiver's rotation transforms
the world-space source--receiver displacement into a local simulator
vector $p_{\mathrm{sim}}$. The model-coordinate vector is
$p=(-z_{\mathrm{sim}},-x_{\mathrm{sim}},y_{\mathrm{sim}})$.
For $p=(x,y,z)$, distance, azimuth, and elevation are
\begin{equation}
\begin{aligned}
r&=\|p\|_2,\quad \theta=\operatorname{atan2}(y,x),\\
\phi&=\operatorname{atan2}\!\left(z,\sqrt{x^2+y^2}\right).
\end{aligned}
\label{eq:position_coordinates}
\end{equation}
Angles are reported in degrees: azimuth is positive leftward in
$[-180^\circ,180^\circ]$, and elevation is positive upward in
$[-90^\circ,90^\circ]$.

Each source is annotated at $t_i=i/12$ seconds for $i=0,\ldots,120$.
Annotations include position, azimuth, elevation, distance, path
speed, source presence, speech activity, and dry-speech energy.
Source presence denotes an existing trajectory, including silent
speech intervals. Speech activity is determined from dry-waveform
RMS energy in an approximately 0.25-second window, with a threshold
of the greater of $-50$\,dBFS and 25\,dB below the clip's maximum
local energy. Presence supervises the source-state model;
speech activity determines valid question times.

% 在同一小节完成查询时刻、五类任务和问题示例的说明。
\subsection{Target-Specific Questions}
\label{app:queries}
\label{app:tasks}
\paragraph{Task definitions.}
At each query time, position direction assigns the target's azimuth
to one of eight sectors, while azimuth, elevation, and distance
return the corresponding numerical attributes. The sector index is
\begin{equation}
\operatorname{sector}(\theta)=
\left\lfloor\frac{\operatorname{wrap}_{360}(\theta+22.5^\circ)}{45^\circ}\right\rfloor,
\label{eq:sector_assignment}
\end{equation}
where $\operatorname{wrap}_{360}$ maps angles to $[0^\circ,360^\circ)$.
Indices follow the sector order given above.
Movement direction uses the horizontal component of
$\Delta p=p_t-p_{t-1\mathrm{s}}$ in the same fixed receiver frame.
Displacements of norm at most $10^{-5}$\,m receive the Stationary
label; other displacements use the sector of
$\operatorname{atan2}(\Delta p_y,\Delta p_x)$.

\paragraph{Question construction.}
Questions identify the target voice and specify either a timestamp
or the preceding one-second interval. Answers are calculated from
the annotated target trajectory. Five tasks at each of two query
times yield ten QA pairs per clip, with 4,800 training and 960 test
questions per task. Table~\ref{tab:sonic_task_protocol} shows examples.
The complete question templates and SonicWorld input instruction are presented in Appendix~\ref{app:prompts}.

\begin{table*}[!htbp]
\centering\footnotesize\color{black}
\setlength{\tabcolsep}{2pt}\renewcommand{\arraystretch}{1.0}
\caption{\textbf{Examples of target-specific spatial questions.}
}
\label{tab:sonic_task_protocol}
\begin{tabularx}{\linewidth}{@{}>{\raggedright\arraybackslash}p{0.18\linewidth}Xc@{}}
\toprule
\textbf{Task} & \textbf{Example question} & \textbf{Answer}\\
\midrule
Position direction & At 3.00 s, which direction is the female voice coming from? & Front\\
Azimuth & At 6.00 s, what is the azimuth of the female voice? & $-9.47^\circ$\\
Elevation & At 4.50 s, what is the vertical angle of the male voice? & $+2.25^\circ$\\
Distance & At 7.00 s, how far away is the male voice? & $3.12$\,m\\
Movement direction & 
% Between 1.50 and 2.50 s, which direction did the female source move relative to the listener's fixed orientation? & 
\makecell[l]{
  Between 1.50 and 2.50 s, which direction did the female source move\\
  relative to the listener's fixed orientation?
} &
Front-Right\\
\bottomrule
\end{tabularx}
\end{table*}

% 集中报告标签分布与连续几何分布。
\subsection{Dataset Distributions}
\label{app:dataset_statistics}
Table~\ref{tab:geometry_stats} describes the source geometry across
all annotated source states. At the target query times, moving and
stationary targets each account for 2,400 training queries and
480 test queries. Table~\ref{tab:test_motion_labels} gives the
movement-direction distribution, and
Figure~\ref{fig:geometry_distributions} shows the detailed attributes at the query times.

\begin{table}[!htbp]
\centering
\footnotesize
\setlength{\tabcolsep}{3pt}
\renewcommand{\arraystretch}{1.10}

\caption{\textbf{Source geometry and trajectory statistics.}}
\label{tab:geometry_stats}

\begin{tabularx}{\linewidth}{
@{}
l
>{\raggedright\arraybackslash}X
rr
@{}
}
\toprule
\textbf{Quantity} &
\textbf{Statistic} &
{\textbf{Train}} &
{\textbf{Test}} \\
\midrule

\multirow{4}{*}{\shortstack[l]{Source range\\(m)}}
 & Minimum & 1.50 & 1.50 \\
 & Maximum & 4.00 & 4.00 \\
 & Mean    & 2.85 & 2.82 \\
 & Median  & 2.90 & 2.87 \\

\midrule

\multirow{5}{*}{Elevation ($^\circ$)}
 & Minimum      & -16.76 & -13.45 \\
 & Maximum      & 19.94  & 20.30  \\
 & P25          & 0.81   & 0.83   \\
 & P50 (median) & 0.99   & 1.01   \\
 & P75          & 1.32   & 1.35   \\

\midrule

\multicolumn{2}{@{}l}{Mean moving-path length (m)}
 & 2.74 & 2.74 \\

\multicolumn{2}{@{}l}{Mean moving speed (m/s)}
 & 0.27 & 0.27 \\

\multicolumn{2}{@{}l}{Minimum dual-source separation (m)}
 & 0.80 & 0.82 \\

\bottomrule
\end{tabularx}
\end{table}

\begin{table*}[!htbp]
\centering
\footnotesize
\setlength{\tabcolsep}{4pt}
\renewcommand{\arraystretch}{1.1}
\caption{\textbf{Test movement-direction distribution.}
Counts denote questions in the single- and dual-source subsets.
S denotes Stationary; F/B/L/R denote Front/Back/Left/Right,
with combined letters indicating diagonal directions.
Stationary labels account for half of each subset.}
\label{tab:test_motion_labels}
\label{tab:query_motion}
\begin{tabular*}{\linewidth}{@{\extracolsep{\fill}}lrrrrrrrrrr@{}}
\toprule
\textbf{Subset} & \textbf{S} & \textbf{F} & \textbf{FL}
& \textbf{L} & \textbf{BL} & \textbf{B} & \textbf{BR}
& \textbf{R} & \textbf{FR} & \textbf{Total}\\
\midrule
Single & 160 & 15 & 30 & 14 & 23 & 23 & 21 & 13 & 21 & 320\\
Dual   & 320 & 55 & 34 & 42 & 37 & 46 & 38 & 40 & 28 & 640\\
Total  & 480 & 70 & 64 & 56 & 60 & 69 & 59 & 53 & 49 & 960\\
\bottomrule
\end{tabular*}
\end{table*}

\begingroup
% Top-align this figure if it is placed on a separate float page.
\makeatletter
\setlength{\@fptop}{0pt}
\makeatother
\begin{figure*}[htbp]
\centering
\includegraphics[width=\linewidth]{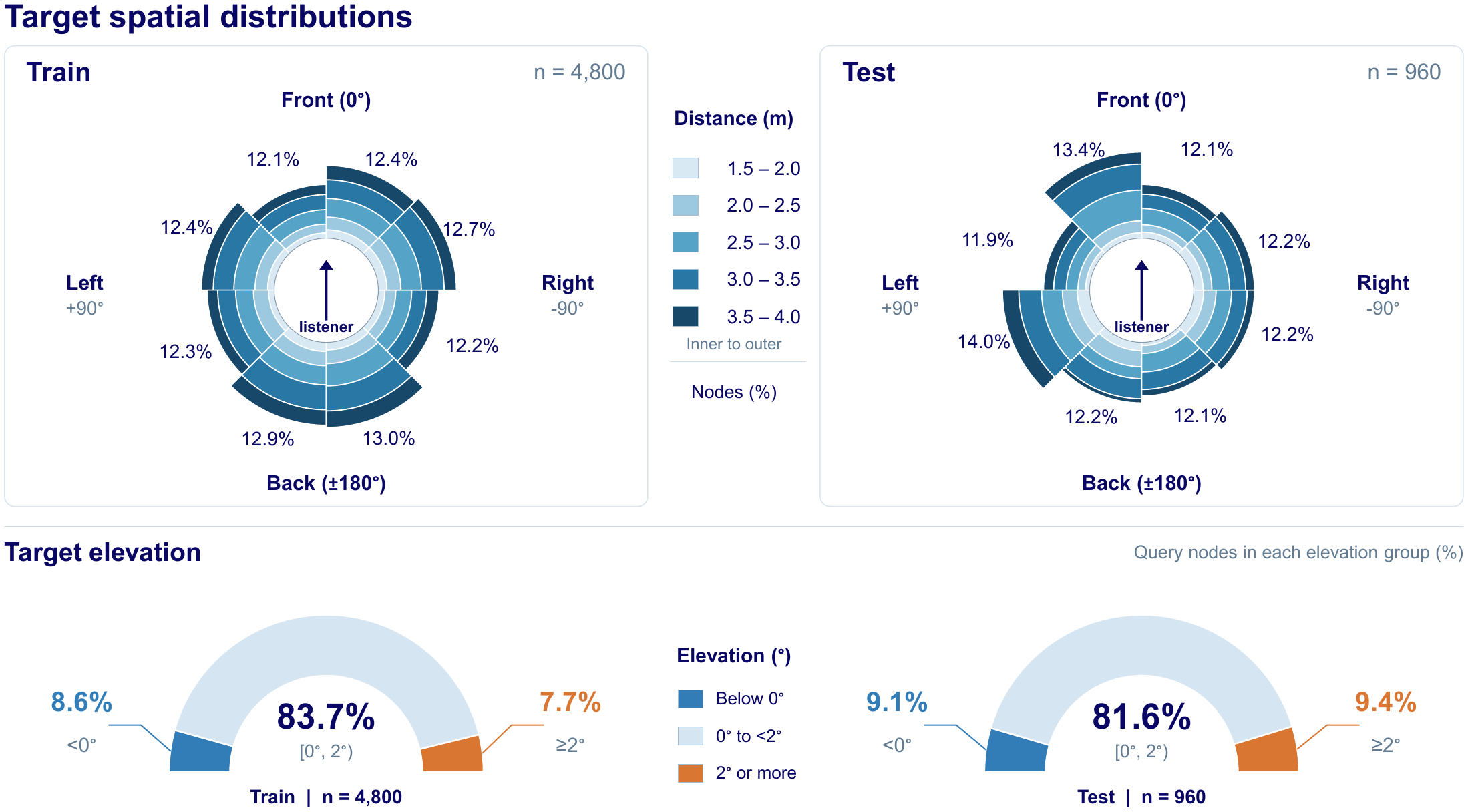}
\caption{\textbf{Target geometry at query times.}
Distributions use 4,800 training and 960 test query times,
expressed as percentages within each split.
The upper panels group azimuth into $45^\circ$ bins starting
at $0^\circ$ and distance into 0.5\,m bands from 1.5 to 4.0\,m.
The lower panels show elevation below $0^\circ$,
in $[0^\circ,2^\circ)$, and at least $2^\circ$.}
\label{fig:geometry_distributions}
\end{figure*}

\endgroup

% \input{tables/dataset_comparison}

% Table~\ref{tab:dataset_comparison} compares Sonic-Bench with
% representative spatial-audio resources.
% Existing datasets span static localization, multimodal grounding,
% and motion reasoning, with several resources including moving-source
% scenarios.
% However, motion is typically covered only in selected subsets or
% task-specific settings, rather than systematically across source
% configurations.
% Sonic-Bench instead covers stationary and moving targets in both
% single- and dual-source scenes, including all target/interferer
% motion combinations.
% Each source is paired with a full-clip trajectory, while every
% question explicitly specifies the target and a timestamp or interval.
% This design supports controlled evaluation of source-specific
% localization, distance estimation, and motion reasoning as acoustic
% scenes evolve over time.

\section{Model Architecture}
\label{app:model}

The Qwen2.5-Omni-7B configuration processes ten-second FOA
recordings at 16\,kHz in channel order $[W,Y,Z,X]$.
Each clip is represented by 121 source-state steps.

\subsection{Source-State Inference}
\label{app:source_state_inference}

\paragraph{Acoustic feature extraction.}
\label{app:features}
The physical branch applies a 400-point short-time Fourier
transform with a 400-sample Hann window and a 160-sample hop.
For directional channel $d\in\{X,Y,Z\}$, $F_d$ and $F_W$
denote its spectrum and the omnidirectional spectrum at the same
time--frequency bin. The cross-spectrum and intensity components are
\begin{equation}
\begin{aligned}
C_d&=F_W^*F_d,\\
A_d&=\operatorname{clip}\!\left(
\frac{\operatorname{Re}C_d}{|F_W|^2},-10,10
\right),\\
R_d&=\operatorname{clip}\!\left(
\frac{\operatorname{Im}C_d}{|F_W|^2},-10,10
\right).
\end{aligned}
\label{eq:physical_intensity}
\end{equation}
Coherence uses averaging over seven adjacent analysis frames,
denoted by $\mathcal A$:
\begin{equation}
\gamma_d=
\frac{|\mathcal A(C_d)|}
{\sqrt{\mathcal A(|F_W|^2)\mathcal A(|F_d|^2)}}.
\label{eq:physical_coherence}
\end{equation}
A fixed 64-band Mel filterbank maps four $\log(1+|F_c|)$
spectra and three components each of active intensity, reactive
intensity, and coherence to thirteen feature channels.
Coherence is clipped to $[0,1]$ before and after Mel projection;
numerical normalizations use $\epsilon=10^{-8}$.
Resampling to 121 steps and a convolutional network
($13\to96\to192\to192$) produce 64 physical tokens of
width $d_e=192$ per step.
All kernels are $3\times3$ with unit stride and padding one;
the first two layers use eight-group normalization and GELU.

A shared 1D CNN separately encodes each waveform channel
(Table~\ref{tab:app_waveform_cnn}). Following Neural IV
\citep{jaeger}, elementwise products of encoded $W$ with each
of $X,Y,Z$ are concatenated into 576 features.
A two-layer mapping ($576\to384\to192$, with GELU)
combines these interactions. Temporal interpolation and a learned
branch embedding produce one neural token per step.

\begin{table}[!htbp]
\centering\footnotesize
\setlength{\tabcolsep}{5pt}
\renewcommand{\arraystretch}{1.08}
\caption{\textbf{Shared waveform encoder.} Each convolution uses
one-group normalization and GELU.}
\label{tab:app_waveform_cnn}
\begin{tabularx}{\linewidth}{@{}Xccc@{}}
\toprule
\textbf{Layer} & \textbf{Channels} & \textbf{Kernel} & \textbf{Stride}\\
\midrule
1 & $1\to48$ & 10 & 5\\
2 & $48\to96$ & 7 & 4\\
3 & $96\to128$ & 5 & 4\\
4 & $128\to192$ & 5 & 4\\
\bottomrule
\end{tabularx}
\end{table}

A frozen BEATs encoder \citep{beats} extracts patch-level semantic features from the omnidirectional $W$ channel.
Layer normalization and a linear projection map them from 768
to 192 dimensions. A learned query pools the patches through
four-head attention and output normalization to obtain one
clip-level semantic token.
At each step, this shared token joins the 64 physical tokens
and one neural token to form $E_t\in\mathbb{R}^{66\times192}$.

\paragraph{Recurrent state estimation and refinement.}
Two source slots and one background slot maintain 256-dimensional
states, starting from learned initial vectors.
At each step, a shared MLP ($256\to512\to256$, with GELU)
predicts a state change. Residual addition and layer normalization
produce the prior $\bar z_t$.
Linear projections map priors to queries and acoustic tokens to
keys and values, all of width 256.
Following Slot Attention \citep{slotattention}, scaled dot-product
scores undergo softmax across slots, followed by normalization
over tokens within each slot.
The weighted value average forms the observation $o_t$.
A 256-dimensional GRU cell \citep{cho2014gru} combines $o_t$
with $\bar z_t$, followed by layer normalization, to obtain
$z_t$ for the next step. The two source sequences continue
to refinement.

% State refinement and geometric prediction.
A shared one-layer bidirectional GRU processes each source
sequence, with 256 hidden units per direction.
Its forward and backward outputs are concatenated and projected
to temporal context $c_t\in\mathbb{R}^{256}$.
A linear layer and sigmoid compute a gate $g_t$ from $[z_t;c_t]$.
The refined state is
$h_t=\operatorname{LayerNorm}(z_t+g_t\odot c_t)$.
Four linear heads predict unit direction $\hat u_t$, log-distance
$\hat d_t$, source presence, and positional log-variance from $h_t$.
Direction is unit-normalized, and log-distance is clipped to
$[\log0.05,\log20]$ before exponentiation. The receiver-centered
position is $\hat p_t=\exp(\hat d_t)\hat u_t$.
Positional log-variance is clipped to $[-6,4]$.

% Source-state supervision.
Shared prediction heads supervise refined trajectories and
transition-prior geometry.
For masked-state prediction, the recurrent module processes
masked sequences in both temporal directions, and the heads
decode the center states before bidirectional refinement.
The objectives are defined in Appendix~\ref{app:training}.

\subsection{Spatial Language Modeling}
\label{app:spatial_language_modeling}

\paragraph{Spatial projector.}
\label{app:projector}
\label{app:anchors}
For each source and time step, the projector concatenates the
prior $\bar z_t$, observation $o_t$, recurrent state $z_t$,
refined state $h_t$, and difference $o_t-\bar z_t$ with the
additional features in Table~\ref{tab:projector_input_fields}.
The direction anchor pools active-intensity directions using
the source-assignment weights.
A frame MLP maps the 1,295-dimensional input to 512 dimensions.
Adding a learned source-slot embedding and sinusoidal time
encoding forms the frame sequence used by both aggregation branches.

\begin{table}[!htbp]
\centering\footnotesize
\setlength{\tabcolsep}{6pt}
\renewcommand{\arraystretch}{1.12}
\caption{\textbf{Projector input per source and time step.}}
\label{tab:projector_input_fields}
\begin{tabularx}{\linewidth}{@{}>{\raggedright\arraybackslash}Xr@{}}
\toprule
\textbf{Input} & \textbf{Dimensions}\\
\midrule
Prior, observation, recurrent state, refined state,
observation minus prior & $5\times256$\\
\addlinespace[4pt]
Direction anchor; predicted direction and distance & $3+3+1$\\
\addlinespace[4pt]
Source-presence probability; positional log-variance & $1+1$\\
\addlinespace[4pt]
Auxiliary features & 6\\
\midrule
\textbf{Total} & \textbf{1,295}\\
\bottomrule
\end{tabularx}
\end{table}

The local branch concatenates non-overlapping groups of four
consecutive frame features and maps each group to one token,
producing 31 local
tokens per source.
In parallel, the summary branch uses two learned queries,
each augmented with the source-slot embedding, to attend to
all 121 frame features through a two-layer, eight-head
Transformer decoder with pre-layer normalization and a
2,048-dimensional feed-forward network using GELU.
Its outputs are two source-summary tokens.
The query parameters and decoder weights are shared across sources.

Both branches produce 512-dimensional tokens.
Their outputs are concatenated across the two sources, giving
66 spatial tokens. An output MLP maps these tokens to the
language embedding width $d_\ell=3584$.
The language model reads them alongside native $W$-channel audio
tokens and the question tokens.
Table~\ref{tab:app_projector_layers} gives the MLP configurations.

\begin{table}[!htbp]
\centering\footnotesize
\setlength{\tabcolsep}{5pt}
\renewcommand{\arraystretch}{1.08}
\caption{\textbf{Projector MLPs.} Each uses layer normalization
at its input and output, with SiLU between the two linear layers.}
\label{tab:app_projector_layers}
\begin{tabularx}{\linewidth}{@{}Xc@{}}
\toprule
\textbf{Module} & \textbf{Linear widths}\\
\midrule
Frame encoder & $1295\to512\to512$\\
Local temporal compressor & $2048\to2048\to512$\\
Language-space projector & $512\to2048\to3584$\\
\bottomrule
\end{tabularx}
\end{table}

\section{Training Objectives and Optimization}
\label{app:training}
\label{app:objectives}

Algorithm~\ref{alg:sonicworld_training} summarizes source-state pretraining and the two
LLM fine-tuning stages.
The trainable source-state parameters $\theta_S$ include the
physical and neural feature branches, semantic projection, recurrent slots, transition MLP, bidirectional
refinement, and prediction heads.
The pretrained BEATs encoder remains frozen.

\begin{algorithm*}[!tp]
% \small
\caption{Training of SonicWorld}
\label{alg:sonicworld_training}
\begin{algorithmic}[1]
\Statex \textbf{Input:} Source-trajectory data $\mathcal D_S$ and spatial QA data $\mathcal D_Q$.
\Statex \textbf{Parameters:} source-state model $\theta_S$; projector $\theta_P$; alignment heads $\theta_A$; LoRA $\theta_L$.
\Statex \textbf{Source-state pretraining}
\State Freeze pretrained BEATs and set $V$ to $4$. $X$ denotes FOA audio and $Y$ the corresponding ground-truth source trajectories.
\For{each minibatch $(X,Y)\in\mathcal D_S$}
  \State $E\gets\operatorname{AcousticFeatures}(X;\theta_S)$
  \State $\widehat Y\gets\operatorname{SourceStates}(E;\theta_S)$
  \State $\pi\gets\operatorname{Hungarian}(\widehat Y,Y)$ 
  % \Comment{one assignment per clip}
  % \State Compute $\mathcal L_{\mathrm{traj}}$ with $\pi$. 
  \State $\mathcal{L}_{\mathrm{traj}} = \operatorname{TrajectoryLoss}(\widehat{Y}, Y; \pi)$
  \Comment{Eq.~\ref{eq:trajectory_loss}}
  \State $\mathcal L_{\mathrm{mask}}=\operatorname{MaskedLoss}(E,Y;\theta_S,V)$ \Comment{Appendix~\ref{app:masking}}
 \State $\mathcal{L}_{\mathrm{state}} = \mathcal{L}_{\mathrm{traj}} + \lambda_{\mathrm{mask}}\mathcal{L}_{\mathrm{mask}}$ 
 % \Comment{Eq.~\ref{eq:state_loss}} 
  \State $\theta_S\gets\operatorname{AdamW}(\theta_S,\nabla_{\theta_S}\mathcal L_{\mathrm{state}})$
\EndFor

\Statex \textbf{Stage I: projector alignment}

\State Freeze $\theta_S$ and original LLM weights. Set $\lambda_P$ to $0.5$; freeze $\theta_L$.
\State Let $q$ denote the question text, $y$ the ground-truth answer, $H$ the source-state features, $G$ the geometric predictions of the frozen source-state model, $Z$ the spatial tokens for the LLM.
\For{each minibatch $(X,q,y)\in\mathcal D_Q$}
  \State $(H,G)\gets\operatorname{SourceFeatures}(X;\theta_S)$ \label{alg:language_forward_start}
 \State $(Z,\widehat G)\gets\operatorname{SpatialProjector}(H,G;\theta_P,\theta_A)$  \Comment{Appendix~\ref{app:projector}}
  \State $\mathcal L_{\mathrm{QA}}=\operatorname{CrossEntropy}(\operatorname{LLM}_{\theta_L}(X,q,Z),y)$ \Comment{Appendix~\ref{app:language_training}}
  \State $\mathcal L_{\mathrm{align}}=\operatorname{AlignLoss}(\widehat G,G)$ \Comment{Eq.~\ref{eq:app_alignment_loss}}
  \State $\mathcal{L}_{\mathrm{LLM}} = \mathcal{L}_{\mathrm{QA}} + \lambda_P\mathcal{L}_{\mathrm{align}}$ \label{alg:language_forward_end}
  % \Comment{Eq.~\ref{eq:lm}}
  \State $(\theta_P,\theta_A)\gets\operatorname{AdamW}((\theta_P,\theta_A),\nabla_{(\theta_P,\theta_A)}\mathcal L_{\mathrm{LLM}})$
\EndFor
\Statex \textbf{Stage II: projector and LoRA adaptation}
\State Set $\lambda_P$ to $0.25$ and enable $\theta_L$ updates.
\For{each minibatch $(X,q,y)\in\mathcal D_Q$}
  \State Repeat lines~\ref{alg:language_forward_start}--\ref{alg:language_forward_end}.
  \State $(\theta_P,\theta_A,\theta_L)\gets\operatorname{AdamW}((\theta_P,\theta_A,\theta_L),\nabla_{(\theta_P,\theta_A,\theta_L)}\mathcal L_{\mathrm{LLM}})$
\EndFor
\State \Return $\theta_S,\theta_P,\theta_A,\theta_L$.
\end{algorithmic}
% \smallskip
% \noindent $\widehat Y$ collects supervised source predictions; $H,G$ are detached source features (prior, observation, recurrent, and refined) and geometric targets; $Z,\widehat G$ are spatial tokens and alignment-head predictions.
% \textsc{MaskedCenterLoss} averages eight independently matched center losses from four masked sequences in both temporal orders, using shared heads before BiGRU refinement.
\end{algorithm*}

\begin{table}[!htbp]
\centering\footnotesize\color{black}
\setlength{\tabcolsep}{5pt}\renewcommand{\arraystretch}{1.12}
\caption{\textbf{Loss coefficients.}}
\label{tab:training_loss_weights}
\begin{tabularx}{\linewidth}{@{}>{\raggedright\arraybackslash}p{0.29\linewidth}>{\raggedright\arraybackslash}X@{}}
\toprule
\textbf{Objective} & \textbf{Weights}\\
\midrule
Trajectory supervision & Direction 1; log-distance 1; presence 1; uncertainty 0.05; prior 0.20\\
\addlinespace[4pt]
Masked center prediction & Direction 1; log-distance 0.3; presence 0.2\\
\addlinespace[4pt]
Projector training & Answer cross-entropy 1; geometric alignment 0.5\\
\addlinespace[4pt]
Projector + LoRA & Answer cross-entropy 1; geometric alignment 0.25\\
\bottomrule
\end{tabularx}
\end{table}

% 损失定义集中在此，架构附录不再重复优化目标和训练系数。
\subsection{Trajectory Matching and Supervision}
\label{app:trajectory_supervision}
Let $p$ denote a ground-truth position, $r=\|p\|_2$ its distance,
$u=p/r$ its unit direction, and $a$ the source-presence label.
For predicted slot $k$ and annotated source $s$, the matching cost is
\begin{equation}
\begin{aligned}
C_{k,s}={}&\overline{1-\hat u_k^\top u_s}
+\overline{\operatorname{SL1}(\hat d_k-\log r_s)}\\
&+0.25\,\operatorname{BCE}(\hat a_k,a_s).
\end{aligned}
\label{eq:app_matching_cost}
\end{equation}
The bars average geometry errors over frames where the source is
present; binary cross-entropy averages over the whole clip.
One Hungarian assignment per clip associates slots with trajectories \citep{kuhn1955hungarian}.
Unmatched slots receive zero-presence targets.

For matched present-source frames indexed by $\mathcal A$, let
$d_i=\log r_i$ and let $s_i$ denote predicted positional log-variance \citep{kendall2017uncertainties}.
Trajectory supervision comprises
\begingroup
\allowdisplaybreaks[4]
\setlength{\abovedisplayskip}{6pt}
\setlength{\abovedisplayshortskip}{6pt}
\setlength{\belowdisplayskip}{6pt}
\setlength{\belowdisplayshortskip}{6pt}
\begin{align}
\mathcal L_{\mathrm{dir}}&=\frac{1}{|\mathcal A|}
\sum_{i\in\mathcal A}(1-\hat u_i^\top u_i),
\label{eq:app_direction_loss}\\
\mathcal L_{\mathrm{dist}}&=\frac{1}{|\mathcal A|}
\sum_{i\in\mathcal A}\operatorname{SL1}(\hat d_i-d_i),
\label{eq:app_distance_loss}\\
\mathcal L_{\mathrm{presence}}&=\frac{1}{BTK}
\sum_{b,t,k}\begin{aligned}[t]
&\operatorname{BCE}(\hat{a}_{btk},a_{btk}),
\end{aligned}
\label{eq:app_presence_loss}\\
\mathcal L_{\mathrm{uncertainty}}&=\frac{1}{2|\mathcal A|}
\sum_{i\in\mathcal A}\left[e^{-s_i}\|\hat p_i-p_i\|_2^2+s_i\right],
\label{eq:app_uncertainty_loss}\\
\mathcal L_{\mathrm{prior}}&=\frac{1}{2|\mathcal A|}
\sum_{i\in\mathcal A}\Bigl[1-(\hat u_i^{\mathrm{prior}})^\top u_i
\nonumber\\*
&\qquad+\operatorname{SL1}(\hat d_i^{\mathrm{prior}}-d_i)\Bigr].
\label{eq:app_prior_loss}
\end{align}
\endgroup
Smooth-L1 is $\operatorname{SL1}(x)=x^2/2$ for $|x|<1$ and
$|x|-1/2$ otherwise. BCE denotes binary cross-entropy for source-presence prediction.
The five terms are combined into $\mathcal L_{\mathrm{traj}}$
with the coefficients in Table~\ref{tab:training_loss_weights}.

% 遮掩采样与中心预测只定义一次；不写缓存、随机种子哈希或张量索引。
\subsection{Masked State Prediction}
\label{app:masking}
We generate $V=4$ masked feature sequences per clip.
For each sequence, we sample an interval length uniformly
from five to ten state steps and select its start with a
six-step boundary margin. All tokens within the interval
are replaced with a learned mask embedding.

The shared recurrent filter processes each sequence $v$
in both temporal directions,
$\delta\in\{\rightarrow,\leftarrow\}$.
At the masked interval's center, shared heads predict direction,
log-distance, and presence from the recurrent states before
bidirectional refinement. These predictions are matched to
the center annotations independently for each $(v,\delta)$.
We average the resulting center-prediction losses
$\ell_{v,\delta}$:
\begin{equation}
\begin{aligned}
\mathcal L_{\mathrm{mask}}
&=\frac{1}{2V}\sum_{v=1}^{V}
\sum_{\delta\in\{\rightarrow,\leftarrow\}}\ell_{v,\delta},\\
\ell_{v,\delta}
&=\mathcal L_{\mathrm{dir}}^{v,\delta}
+0.3\mathcal L_{\mathrm{dist}}^{v,\delta}
+0.2\mathcal L_{\mathrm{presence}}^{v,\delta}.
\end{aligned}
\label{eq:mask_loss}
\end{equation}
% where each component loss is evaluated at the center of masked
% sequence $v$ in temporal order $\delta$.

% Source-state pretraining combines
% $\mathcal L_{\mathrm{traj}}+\lambda_{\mathrm{mask}}\mathcal L_{\mathrm{mask}}$.
% The mask coefficient is zero in epochs 1--2, increases through
% 0.05, 0.10, 0.15, and 0.20 in epochs 3--6, and remains at 0.20.

% 与轨迹监督区分：语言阶段用冻结模型预测来对齐 projector。
\subsection{Two-stage LLM Fine-Tuning}
\label{app:language_training}
After source-state pretraining, we train the spatial
projector and its alignment heads in Stage~I. In Stage~II, we additionally train LoRA parameters in the LLM's
attention layers. The source-state model, native audio encoder, and original
LLM weights remain frozen in both stages.
Both stages optimize cross-entropy on answer tokens together with
geometric alignment.
Alignment heads predict direction, log-distance, and presence
from projector frame features after source and time encoding
and before temporal compression.
Table~\ref{tab:app_representation_shapes} lists the intermediate
representations.

$P$ and $*$ denote alignment-head predictions
and geometric predictions from the frozen source-state
model, respectively. We set $w_i=\max(\hat a_i^*,0.05)$. The alignment loss is
\begin{equation}
\begin{aligned}
\mathcal L_{\mathrm{align}}
={}&\frac{\sum_i w_i(1-(\hat u_i^P)^\top\hat u_i^*)}{\sum_i w_i}\\
&+0.5\,\operatorname{mean}_i\operatorname{SL1}(\hat d_i^P-\hat d_i^*)
\\
&+0.25\,\operatorname{mean}_i\operatorname{BCE}(\hat a_i^{P},\hat a_i^*).
\end{aligned}
\label{eq:app_alignment_loss}
\end{equation}
Here, $i$ indexes clip, time-step, and source-slot combinations,
and $\hat a_i^\ast$ is the target source-presence probability.
The overall objective is
$\mathcal L_{\mathrm{QA}}+\lambda_P\mathcal L_{\mathrm{align}}$,
with stage-specific weights in Table~\ref{tab:training_loss_weights}.

\begin{table}[!htbp]
\centering\footnotesize
\setlength{\tabcolsep}{5pt}
\renewcommand{\arraystretch}{1.08}
\caption{\textbf{Model representations from input to output.}
$B$ denotes batch size. Slot--token weights include two source slots
and one background slot; subsequent source representations retain
only the two source slots.}
\label{tab:app_representation_shapes}
\begin{tabularx}{\linewidth}{@{}Xc@{}}
\toprule
\textbf{Representation} & \textbf{Shape}\\
\midrule
FOA waveform, $[W,Y,Z,X]$ & $[B,4,160000]$\\
Physical input features & $[B,13,64,121]$\\
Physical tokens & $[B,121,64,192]$\\
Neural tokens / semantic tokens & $[B,121,1,192]$ \\
Combined acoustic tokens & $[B,121,66,192]$\\
Slot--token weights & $[B,3,66]$\\
Retained source assignments & $[B,121,2,66]$\\
Prior / observation / recurrent / refined states
& $[B,121,2,256]$ \\
Direction / Cartesian position & $[B,121,2,3]$ \\
Distance / presence / log-variance & $[B,121,2]$ \\
Projector frame features & $[B,121,2,512]$\\
Local tokens & $[B,2,31,512]$\\
Summary tokens & $[B,2,2,512]$\\
Spatial tokens for the LLM & $[B,66,3584]$\\
\bottomrule
\end{tabularx}
\end{table}

% 所有实际使用的超参数集中成表，不列未启用实验分支。
\subsection{Optimization Settings}
\label{app:optimization}
The main experiments use the 2,400-clip training split and its
24,000 questions. All stages use AdamW \citep{loshchilov2019adamw}, weight decay 0.01,
cosine learning-rate decay, 5\% learning-rate warmup, bfloat16,
gradient clipping at norm 1, and a micro-batch size one.
Table~\ref{tab:training_stages} lists the stage-specific settings.

\begin{table*}[!htbp]
\centering\footnotesize\color{black}
\setlength{\tabcolsep}{2pt}\renewcommand{\arraystretch}{1.0}
\caption{\textbf{SonicWorld training stages.} Updates count optimizer steps.}
\label{tab:training_stages}
\begin{tabularx}{\linewidth}{@{}>{\raggedright\arraybackslash}p{0.16\linewidth}cccX@{}}
\toprule
\textbf{Stage} & \textbf{Epochs} & \textbf{Accum.} & \textbf{Updates} & \textbf{Trainable modules and peak learning rate}\\
\midrule
Source states & 25 & 2 & 30,000 & Physical and neural branches, semantic pooling, recurrent filter, bidirectional GRU, state heads: $10^{-4}$\\
Projector & 5 & 8 & 15,000 & Projector and alignment heads: $10^{-4}$\\
Projector + LoRA & 3 & 8 & 9,000 & Projector : $3\times10^{-5}$; LoRA: $5\times10^{-5}$\\
\midrule
\multicolumn{4}{@{}l}{\textbf{GPUs}} & $3\times$ NVIDIA RTX 4090; $6\times$ NVIDIA RTX 3090 \\
\bottomrule
\end{tabularx}
\end{table*}

LoRA modifies the attention query, key, value, and output
projections with rank 16, scaling 32, and dropout 0.05
\citep{hu2021lora}. Each backbone
retains its native frozen audio encoder and uses five projector
epochs followed by three joint projector--LoRA epochs.

% \paragraph{Two-stage fine-tuning.}
% The first stage trains the projector and alignment heads;
% the second also updates LoRA parameters in the LLM attention layers.
% The source-state model, native audio encoder, and original LLM
% weights remain frozen throughout.
% Alignment heads predict direction, log-distance, and presence
% from frame features before temporal compression, using the frozen
% source-state model's predictions as targets.
% Training objectives and settings are detailed in
% Appendix~\ref{app:training}. Table~\ref{tab:app_representation_shapes} summarizes the
% representations along the complete forward path.

\section{Supplementary Results}
\label{app:results}
\begin{table*}[htbp]
\centering\footnotesize\color{black}
\setlength{\tabcolsep}{3pt}
\renewcommand{\arraystretch}{1.10}
\caption{\textbf{Effect of SonicWorld on different backbones on Sonic-Bench.} \textcolor{black}{Values in parentheses indicate changes relative to the corresponding original backbone without SonicWorld.}}
\label{tab:backbones}

\resizebox{\linewidth}{!}{%
\begin{tabular}{@{}l*{5}{r}@{}}
\toprule
Backbone
& Dir.$\uparrow$ & Az.$\downarrow$ & El.$\downarrow$
& Dist.$\downarrow$ & Mot.$\uparrow$\\
\midrule
\multicolumn{6}{c}{\textbf{Single source}}\\
\midrule

Qwen2.5-Omni-7B
& 50.00 ($+37.50$)
& \textbf{30.85} ($-60.33$)
& \textbf{0.91} ($-84.00$)
& 0.56 ($-98.27$)
& 52.50 ($+0.62$)\\

InternOmni
& \textbf{51.88} ($+39.07$)
& 36.49 ($-57.23$)
& 0.92 ($-21.73$)
& 0.61 ($-39.97$)
& \textbf{55.94} ($+47.81$)\\

Phi-4-MM-5.6B
& 48.75 ($+35.31$)
& 32.03 ($-49.18$)
& 0.92 ($-2.37$)
& \textbf{0.53} ($-1.40$)
& 53.44 ($+47.19$)\\

\midrule
\multicolumn{6}{c}{\textbf{Dual source}}\\
\midrule
Qwen2.5-Omni-7B
& 27.03 ($+14.37$)
& \textbf{56.59} ($-31.74$)
& \textbf{0.93} ($-82.65$)
& \textbf{0.60} ($-33.70$)
& \textbf{53.91} ($+5.32$)\\

InternOmni
& 30.16 ($+17.82$)
& 62.49 ($-30.41$)
& 0.93 ($-31.58$)
& 0.62 ($-42.54$)
& 52.19 ($+46.25$)\\

Phi-4-MM-5.6B
& \textbf{30.78} ($+18.12$)
& 60.50 ($-19.40$)
& 0.95 ($-3.18$)
& 0.60 ($-1.66$)
& 52.03 ($+46.72$)\\
\bottomrule
\end{tabular}%
}
\end{table*}
\subsection{Evaluation Protocol and Metrics}
\label{app:benchmark}
% \paragraph{Localization metrics.}
% Position direction uses exact-label accuracy. For numerical answers,
% the errors are the shortest circular azimuth difference,
% absolute elevation difference, and absolute distance difference.
% MAE averages these errors over finite parsed responses.
% Tolerance accuracy counts predictions within $20^\circ$ for azimuth,
% $1^\circ$ for elevation, and 0.5\,m for distance.

% \paragraph{Motion metrics.}
% Movement-direction accuracy (Mot.) requires the exact direction label,
% including Stationary. Moving-target direction accuracy (M-dir) uses
% only questions with nonstationary ground truth.

% \subsection{Complete Results}
% Table~\ref{tab:qa_single} reports single- and dual-source results for
% closed-source models, open-source audio models, spatial baselines,
% and all four SonicWorld language backbones. Each SonicWorld variant
% uses the same frozen source-state model and an independently trained
% projector and LoRA adapter.
% \input{tables/appendix_results_1}

\paragraph{Metrics details.}
We use the metrics defined in Section~\ref{sec:evaluation}.
MAE is computed over finite parsed numerical responses.

\subsection{Effect of SonicWorld on Different Backbones}
\label{app:backbones}

On Sonic-Bench, SonicWorld improves all five metrics in both
source settings \textcolor{black}{relative to the original Qwen2.5-Omni-7B, InternOmni, and Phi-4-MM-5.6B backbones}
(Table~\ref{tab:backbones}).
Direction accuracy gains range from 35.31--39.07 percentage
points for single sources and 14.37--18.12 points for dual sources.

On SLS (Table~\ref{tab:external_backbones}),
SonicWorld increases direction accuracy by 54.42--61.29 percentage
points and reduces azimuth MAE by 67.49--71.21$^\circ$ across
the three matched backbones. On SpatialSceneQA, all three variants
reduce azimuth MAE; elevation MAE decreases for Qwen2.5-Omni-7B
and InternOmni but increases by 2.03$^\circ$ for Phi-4-MM-5.6B.

\begin{table*}[htbp]
\centering\footnotesize\color{black}
\setlength{\tabcolsep}{3pt}
\renewcommand{\arraystretch}{1.10}
\caption{\textbf{Effect of SonicWorld on zero-shot transfer across backbones.} \textcolor{black}{Values in parentheses indicate changes relative to the corresponding original backbone without SonicWorld.}}
\label{tab:external_backbones}

\resizebox{\linewidth}{!}{%
\begin{tabular}{@{}l*{6}{r}@{}}
\toprule
& \multicolumn{2}{c}{SpatialSceneQA}
& \multicolumn{4}{c}{SLS}\\
\cmidrule(lr){2-3}\cmidrule(l){4-7}
Backbone
& Az.$\downarrow$ & El.$\downarrow$
& Dir.$\uparrow$ & Az.$\downarrow$
& El.$\downarrow$ & Dist.$\downarrow$\\
\midrule

Qwen2.5-Omni-7B
& 21.35 ($-39.12$)
& 20.51 ($-59.63$)
& \textbf{73.98} ($+61.29$)
& 19.29 ($-70.97$)
& 7.09 ($-81.38$)
& 0.89 ($-54.16$)\\

InternOmni
& 22.57 ($-57.97$)
& 20.58 ($-57.47$)
& 66.80 ($+54.42$)
& 22.20 ($-67.49$)
& \textbf{7.08} ($-50.51$)
& 1.24 ($-97.59$)\\

Phi-4-MM-5.6B
& \textbf{19.79} ($-42.07$)
& \textbf{20.34} ($+2.03$)
& 73.44 ($+60.83$)
& \textbf{18.22} ($-71.21$)
& 7.95 ($+0.87$)
& \textbf{0.81} ($-1.18$)\\
\bottomrule
\end{tabular}%
}
\end{table*}

\subsection{Localization Tolerance Accuracy}

Table~\ref{tab:threshold_all} shows that SonicWorld achieves
the highest localization tolerance accuracies in both source settings.
Az20 exceeds Spatial-Omni by 22.82 percentage points on
single-source clips and 4.53 points on dual-source clips.
From single to dual sources, SonicWorld's Az20 drops from
54.38\% to 25.16\%, while El1 remains near 81\% and D0.5
stays at 53.75\%, suggesting that source interference particularly
affects precise horizontal localization.
The concentrated elevation distribution
(Table~\ref{tab:geometry_stats}) may partly explain the stable El1.

On El1, SonicWorld exceeds BAT by 0.93 percentage points for
single sources and 0.94 points for dual sources. Its D0.5 accuracy
is 53.75\% in both settings, compared with 52.81\% for Neural-IV on
single-source clips and 50.47\% for Classical-IV on dual-source clips.

\begin{table*}[!htbp]
\centering\footnotesize\color{black}
\setlength{\tabcolsep}{2pt}
\renewcommand{\arraystretch}{1.05}
\caption{\textbf{Localization tolerance accuracies.}
Spatial baselines and SonicWorld use Qwen2.5-Omni-7B.
Az20, El1, and D0.5 denote azimuth, elevation, and distance accuracies within $20^\circ$, $1^\circ$, and 0.5\,m, respectively.
}
% All values are percentages. Az20, El1, and D0.5 use tolerances
% of $20^\circ$, $1^\circ$, and 0.5\,m.
% M-dir measures exact movement-direction accuracy on moving targets.
% Best values in each column are bold.
\label{tab:threshold_single}
\label{tab:threshold_dual}
\label{tab:threshold_all}

\begin{tabularx}{\linewidth}{
@{}l*{6}{>{\centering\arraybackslash}X}@{}
}
\toprule
& \multicolumn{3}{c}{Single source}
& \multicolumn{3}{c}{Dual source}\\
\cmidrule(lr){2-4}\cmidrule(l){5-7}
Model
& Az20$\uparrow$ & El1$\uparrow$ & D0.5$\uparrow$ % & M-dir$\uparrow$
& Az20$\uparrow$ & El1$\uparrow$ & D0.5$\uparrow$ % & M-dir$\uparrow$
\\
\midrule
\multicolumn{7}{c}{\textbf{Closed-source models}}\\
\midrule
Gemini 3 Flash
& 10.00 & 30.31 & 25.94 % & 11.25
& 11.72 & 27.34 & 26.41 % & 10.94
\\
Gemini 3.1 Pro
& 12.19 & 31.88 & 1.88 % & 6.88
& 12.03 & 18.59 & 6.25 % & 6.25
\\
\midrule
\multicolumn{7}{c}{\textbf{Open-source models}}\\
\midrule
Qwen2.5-Omni-7B
& 14.06 & 1.88 & 0.00 % & 3.75
& 11.72 & 1.09 & 0.00 % & 3.75
\\
InternOmni
& 9.06 & 23.75 & 23.44 % & \textbf{16.25}
& 9.06 & 22.34 & 24.69 % & 11.88
\\
Phi-4-MM-5.6B
& 5.00 & 1.25 & 4.69 % & 12.50
& 7.03 & 1.09 & 1.56 % & 10.63
\\
Audio Flamingo 3
& 9.69 & 5.63 & 18.13 % & 15.63
& 10.63 & 6.09 & 11.25 % & 10.31
\\
Kimi-Audio-7B-Instruct
& 11.25 & 43.13 & 9.06 % & 3.13
& 11.72 & 40.16 & 9.84 % & 11.88
\\
\midrule
\multicolumn{7}{c}{\textbf{Spatial baselines}}\\
\midrule
BAT
& 10.31 & 79.38 & 45.63 % & 9.38
& 11.09 & 80.47 & 49.84 % & 11.88
\\
Classical-IV
& 30.63 & 75.94 & 50.94 % & 8.13
& 18.91 & 75.00 & 50.47 % & 10.94
\\
Neural-IV
& 14.69 & 76.56 & 52.81 % & 8.13
& 12.34 & 74.22 & 45.00 % & 11.25
\\
Spatial-Omni
& 31.56 & 67.50 & 30.31 % & 8.75
& 20.63 & 72.34 & 28.13 % & 10.00
\\
\midrule
\rowcolor{SWTableHighlight}
\textbf{\method{}}
& \textbf{54.38} & \textbf{80.31} & \textbf{53.75} % & 11.88
& \textbf{25.16} & \textbf{81.41} & \textbf{53.75} % & \textbf{13.75}
\\
\bottomrule
\end{tabularx}
\end{table*}

\subsection{Zero-Shot Transfer Protocol and Results}
\label{app:external_evaluation}
We evaluate the official test splits of SpatialSceneQA
\citep{jaeger} and Spatial LibriSpeech (SLS)
\citep{spatial_librispeech2023}, retaining recordings of at least
10 seconds and using their first 10 seconds at 16\,kHz.
Table~\ref{tab:external_selection} summarizes their scale and
task coverage: SpatialSceneQA tests single- and dual-source
angular localization, while SLS covers four single-source spatial
attributes. Both assess static localization transfer.

\begin{table}[htbp]
\centering\footnotesize\color{black}
\setlength{\tabcolsep}{2pt}\renewcommand{\arraystretch}{1.0}
\caption{\textbf{Evaluation-set scale and selection.} Sonic-Bench uses its full test set; SpatialSceneQA and SLS use the selected test subsets. Dir., Az., El., Dist., and Mot. denote direction, azimuth, elevation, distance, and movement direction.}
\label{tab:external_selection}
\begin{tabularx}{\linewidth}{@{}>{\raggedright\arraybackslash}p{0.36\linewidth}*{3}{>{\centering\arraybackslash}X}@{}}
\toprule
 & \shortstack{Sonic-\\Bench} & \shortstack{Spatial\\SceneQA} & SLS\\
\midrule
Evaluated clips & 480 & 622 & 8,512\\
Single-source clips & 160 & 247 & 8,512\\
Dual-source clips & 320 & 375 & 0\\
Motion state & \shortstack{Stationary /\\Moving} & Stationary & Stationary\\
\midrule
Questions per evaluated clip & 10 & 2 & 4\\
Single-source questions & 1,600 & 494 & 34,048\\
Dual-source questions & 3,200 & 750 & 0\\
Total questions & 4,800 & 1,244 & 34,048\\
\midrule
Query types & \shortstack{Dir., Az., El.,\\Dist., Mot.} & Az., El. & \shortstack{Dir., Az.,\\El., Dist.}\\
\bottomrule
\end{tabularx}
\end{table}

\begin{table*}[!tp]
\centering\footnotesize\color{black}
\setlength{\tabcolsep}{2pt}
\renewcommand{\arraystretch}{1.05}
\caption{\textbf{Single- and dual-source SpatialSceneQA results.}
Spatial baselines and SonicWorld use Qwen2.5-Omni-7B.
}
% Each clip contributes one azimuth and one elevation question.

\label{tab:external_source_counts}

\begin{tabularx}{\linewidth}{
@{}l*{8}{>{\centering\arraybackslash}X}@{}
}
\toprule
& \multicolumn{4}{c}{Single source}
& \multicolumn{4}{c}{Dual source}\\
\cmidrule(lr){2-5}\cmidrule(l){6-9}
Model
& Az.$\downarrow$ & El.$\downarrow$ & Az20$\uparrow$ & El1$\uparrow$
& Az.$\downarrow$ & El.$\downarrow$ & Az20$\uparrow$ & El1$\uparrow$\\
\midrule
\multicolumn{9}{c}{\textbf{Open-source audio models}}\\
\midrule
Qwen2.5-Omni-7B
& 20.54 & 100.49 & 53.85 & 0.00
& 86.77 & 66.73 & 23.20 & 0.00\\
InternOmni
& 75.35 & 83.90 & 5.26 & 0.00
& 83.89 & 74.26 & 4.53 & 0.00\\
Phi-4-MM-5.6B
& 78.49 & 16.42 & 10.93 & 0.40
& 47.69 & 21.14 & 30.13 & 0.00\\
Audio Flamingo 3
& 117.98 & 27.04 & 14.17 & 0.40
& 145.38 & 22.64 & 3.47 & 0.00\\
Kimi-Audio-7B-Instruct
& 53.70 & 34.07 & 30.77 & 0.81
& 56.48 & 36.28 & 29.33 & 0.00\\
\midrule
\multicolumn{9}{c}{\textbf{Spatial baselines}}\\
\midrule
BAT
& 117.35 & 18.01 & 8.91 & 0.00
& 130.65 & 22.15 & 0.00 & 0.00\\
Classical-IV
& 25.47 & 15.78 & 45.75 & 1.62
& 38.88 & 21.40 & 22.40 & 0.00\\
Neural-IV
& 90.45 & \textbf{14.61} & 22.27 & \textbf{2.83}
& 86.94 & \textbf{21.05} & 20.27 & 0.00\\
Spatial-Omni
& 132.55 & 21.71 & 0.40 & 0.00
& 92.36 & 23.59 & 17.60 & \textbf{0.53}\\
\midrule
\rowcolor{SWTableHighlight}
\textbf{\method{}}
& \textbf{16.28} & 17.84 & \textbf{66.40} & 0.00
& \textbf{24.69} & 22.27 & \textbf{51.73} & 0.00\\
\bottomrule
\end{tabularx}
\end{table*}

\begin{table*}[!htbp]
\centering\footnotesize\color{black}
\setlength{\tabcolsep}{3pt}
\renewcommand{\arraystretch}{1.05}
\caption{\textbf{Results on SLS with official spatial interference.}
Evaluation covers 8,512 clips and 34,048 questions.
Spatial baselines and SonicWorld use Qwen2.5-Omni-7B.
Dir. is accuracy (\%); Az./El. are MAE in degrees
and Dist. is MAE in meters. Best values are bold.}
\label{tab:sls_interference}

\begin{tabularx}{\linewidth}{
@{}l*{4}{>{\centering\arraybackslash}X}@{}
}
\toprule
Model
& Dir.$\uparrow$ & Az.$\downarrow$
& El.$\downarrow$ & Dist.$\downarrow$\\
\midrule
\multicolumn{5}{c}{\textbf{Open-source models}}\\
\midrule
Qwen2.5-Omni-7B
& 12.01 & 90.68 & 14.40 & 19.61\\
InternOmni
& 12.76 & 89.19 & 54.55 & 7.12\\
Phi-4-MM-5.6B
& 12.48 & 90.42 & \textbf{7.04} & 0.96\\
Audio Flamingo 3
& 12.31 & 89.46 & 16.39 & 3.11\\
Kimi-Audio-7B-Instruct
& 11.83 & 90.81 & 8.59 & 2.81\\
\midrule
\multicolumn{5}{c}{\textbf{Spatial baselines}}\\
\midrule
BAT
& 12.50 & 90.06 & 7.07 & \textbf{0.84}\\
Classical-IV
& 13.15 & 77.32 & 8.07 & 1.38\\
Neural-IV
& 12.91 & 90.50 & 8.28 & 1.34\\
Spatial-Omni
& 26.97 & 71.69 & 13.88 & 6.55\\
\midrule
\rowcolor{SWTableHighlight}
\textbf{\textsc{SonicWorld}}
& \textbf{63.35} & \textbf{26.74} & 7.08 & 1.16\\
\bottomrule
\end{tabularx}
\end{table*}

% \paragraph{Question construction and input.}
For SpatialSceneQA, we split each joint azimuth/elevation query
into two scalar questions, preserving the target and labels.
For SLS, we construct direction, azimuth, elevation, and distance
questions from speech-source annotations, converting angles to
degrees and assigning direction labels via
Equation~\ref{eq:sector_assignment}.
SpatialSceneQA channels are mapped as
$(W,Y,Z,X)=(c_0,-c_3,c_1,-c_2)$; SLS already uses this order.
Spatial encoders receive all four channels, while native audio
encoders and direct-audio baselines receive the same crop's
$W$ channel. Scoring uses the Sonic-Bench answer parser.

% \paragraph{Results.}
Table~\ref{tab:external_source_counts} reports SpatialSceneQA
results with Qwen2.5-Omni-7B as SonicWorld's backbone.
On dual-source clips, adaptation reduces azimuth MAE from
86.77$^\circ$ to 24.69$^\circ$ and raises Az20 from 23.20\%
to 51.73\%. On single-source clips, furthermore, azimuth MAE
decreases from 20.54$^\circ$ to 16.28$^\circ$, showing that the azimuth
transfer gains extend to both single- and dual-source settings.
Aggregate external results appear in
Table~\ref{tab:zeroshot_results}; backbone comparisons are
provided in Appendix~\ref{app:backbones}.

\subsection{SLS with Spatial Interference}
\label{app:sls_interference}

We evaluate robustness to spatial interference using the official
paired speech and noise recordings from SLS
\citep{spatial_librispeech2023}.
Answers follow the speech-source annotations.
As shown in Table~\ref{tab:sls_interference}, SonicWorld improves direction accuracy from 26.97\% to 63.35\% and reduces azimuth MAE from $71.69^\circ$ to $26.74^\circ$ compared with Spatial-Omni.
Its elevation MAE (7.08$^\circ$) is close to the best result
(7.04$^\circ$), while distance MAE (1.16\,m) remains above
BAT's 0.84\,m.
These results show a clear advantage in bearing estimation
under spatial interference, with more limited benefits
for elevation and distance.

\FloatBarrier
\raggedbottom
\makeatletter
\setlength{\@fptop}{0pt}
\makeatother

\section{Case Studies}
\label{app:case_studies}

In this section, we examine SonicWorld's motion predictions and decoded source
trajectories in single- and dual-source scenes with moving
and stationary speakers.

\subsection{Prediction Comparison}
\label{app:case_qa}

\begin{figure*}[!htbp]
  \centering
  \includegraphics[width=\linewidth]{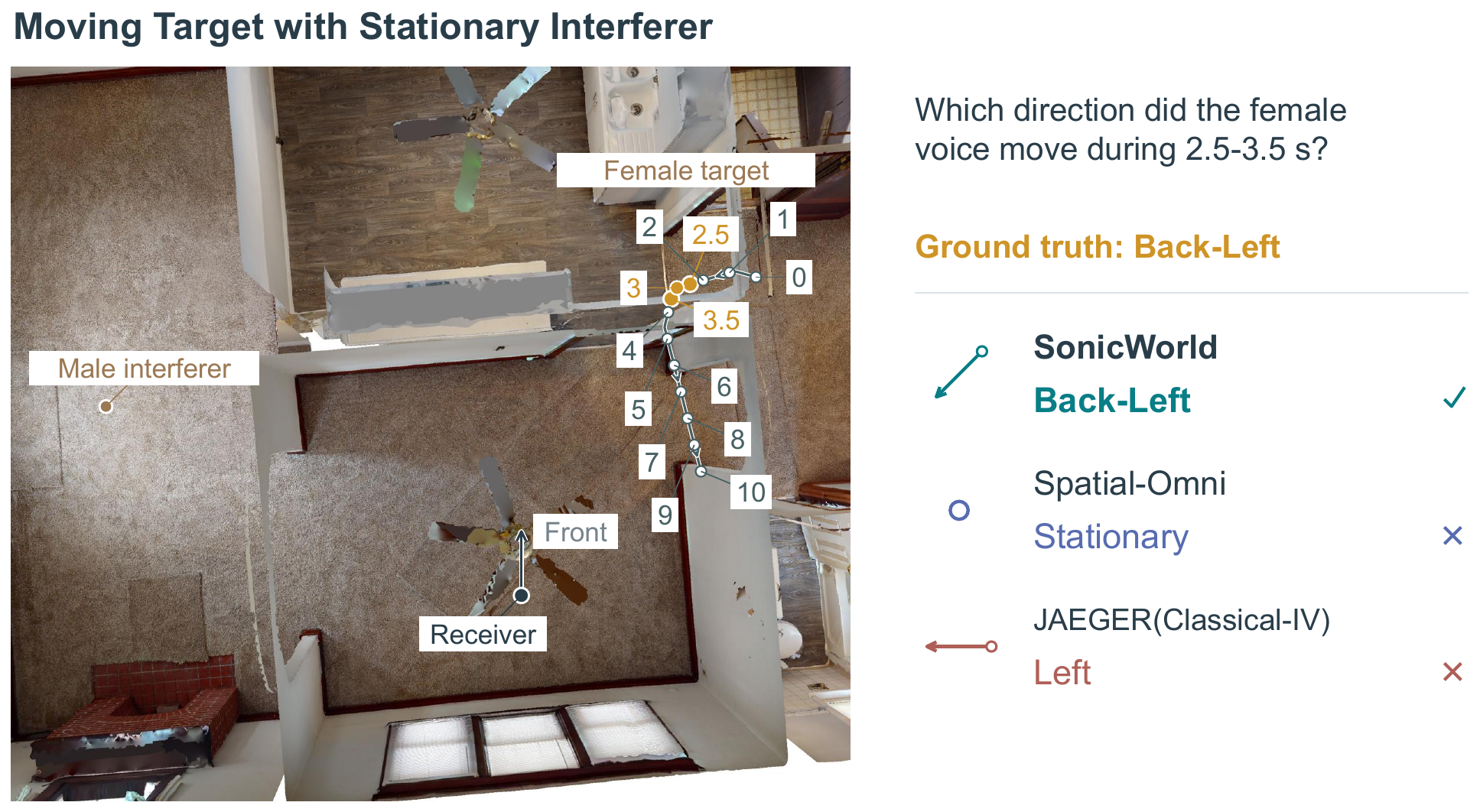}
  \caption{\textbf{Movement predictions with a stationary interferer.}}
  \label{fig:case_study_turns_interferer}
\end{figure*}

\begin{figure*}[!htbp]
  \centering
  \includegraphics[width=\linewidth]{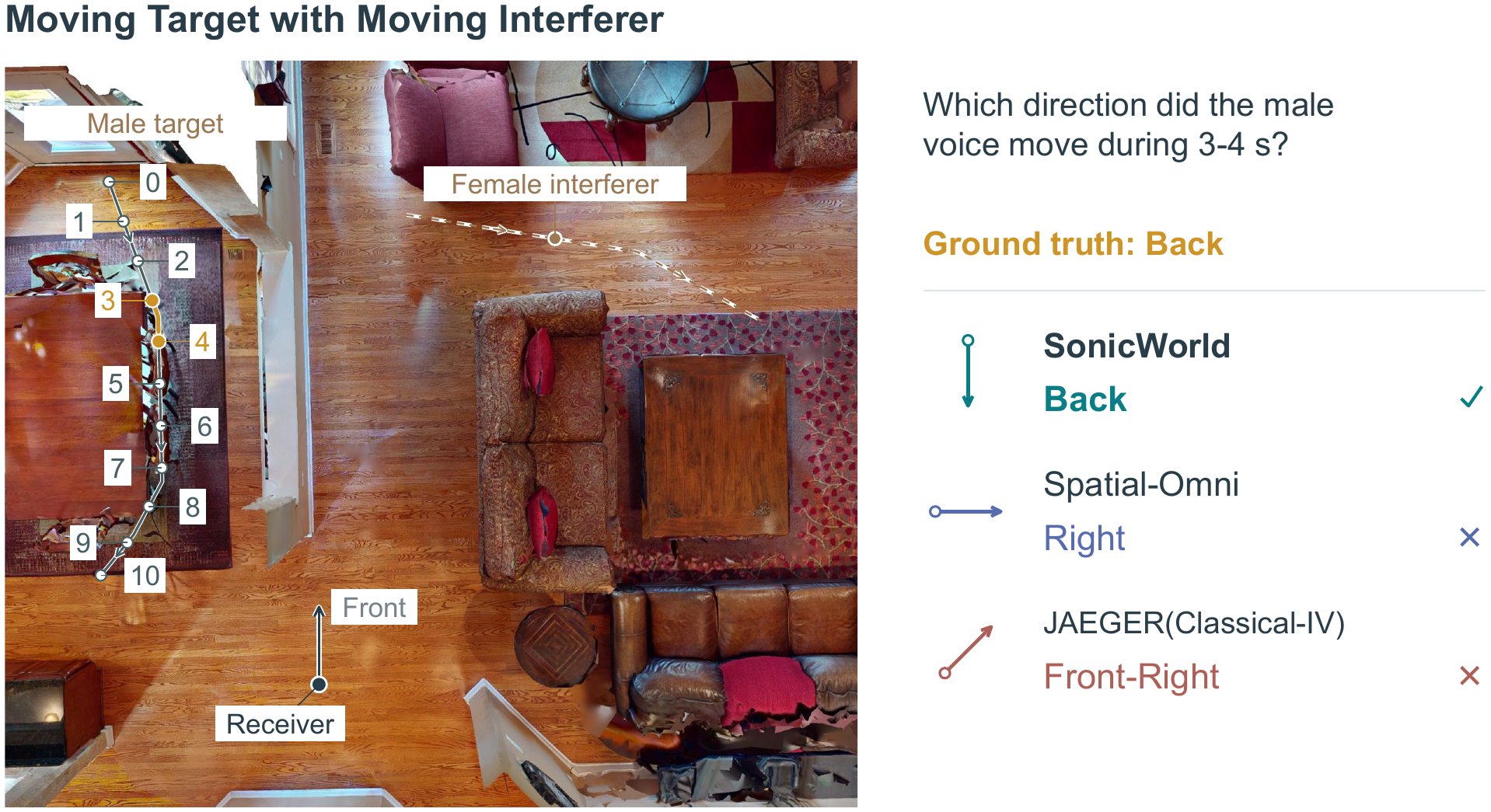}
  \caption{\textbf{Movement predictions for a male target
  with a moving interferer.}}
  \label{fig:case_study_interference}
\end{figure*}

% \paragraph{Motion through a turn.}
% \paragraph{Target motion under interference.}
Figures~\ref{fig:case_study_turns_interferer}
and~\ref{fig:case_study_interference} illustrate target-specific
motion reasoning through turns and under interference.
SonicWorld correctly predicts Back-Left with a stationary
interferer and Back for a male target with a moving interferer,
while Spatial-Omni and JAEGER (Classical-IV) return incorrect
directions or Stationary.

\subsection{Persistent Source-State Analysis}
\label{app:case_states}

We next examine the spatial representations underlying
motion reasoning. Each source-state sequence is decoded
into receiver-centered coordinates and distances at
121 time steps over 0--10\,s and compared with ground truth.
For multi-source clips, a single sequence-level assignment
matches the source slots to the annotated trajectories.

\begin{figure*}[!tp]
  \centering
  \includegraphics[width=\linewidth,height=0.4\textheight,keepaspectratio]{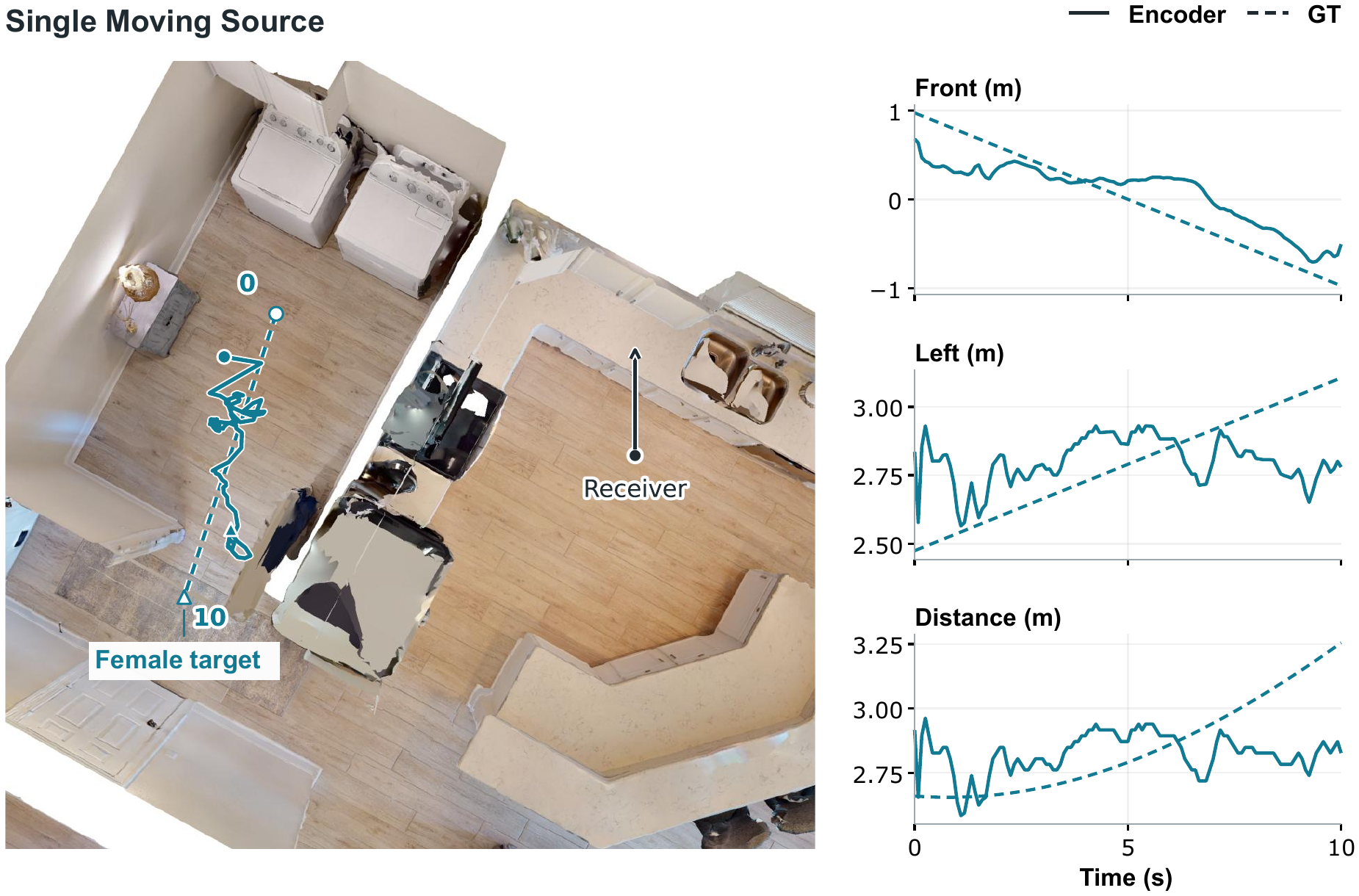}
  \caption{\textbf{Decoded states of a single moving source.}
  Solid lines show decoded trajectories and dashed lines show
  ground truth. The scene view shows the spatial paths;
  the time series show receiver-centered front and left
  coordinates and source distance. Circles and triangles
  mark trajectory starts and ends.}
  \label{fig:case_state_single}
\end{figure*}

\paragraph{Following a moving source.}
In Figure~\ref{fig:case_state_single}, the predicted front
coordinate decreases from positive to negative values,
capturing the source's progression from the receiver's
front side to its rear side. The decoded trajectory has
an average displacement error of 0.342\,m.

\begin{figure*}[!tp]
  \centering
  \includegraphics[width=\linewidth,height=0.4\textheight,keepaspectratio]{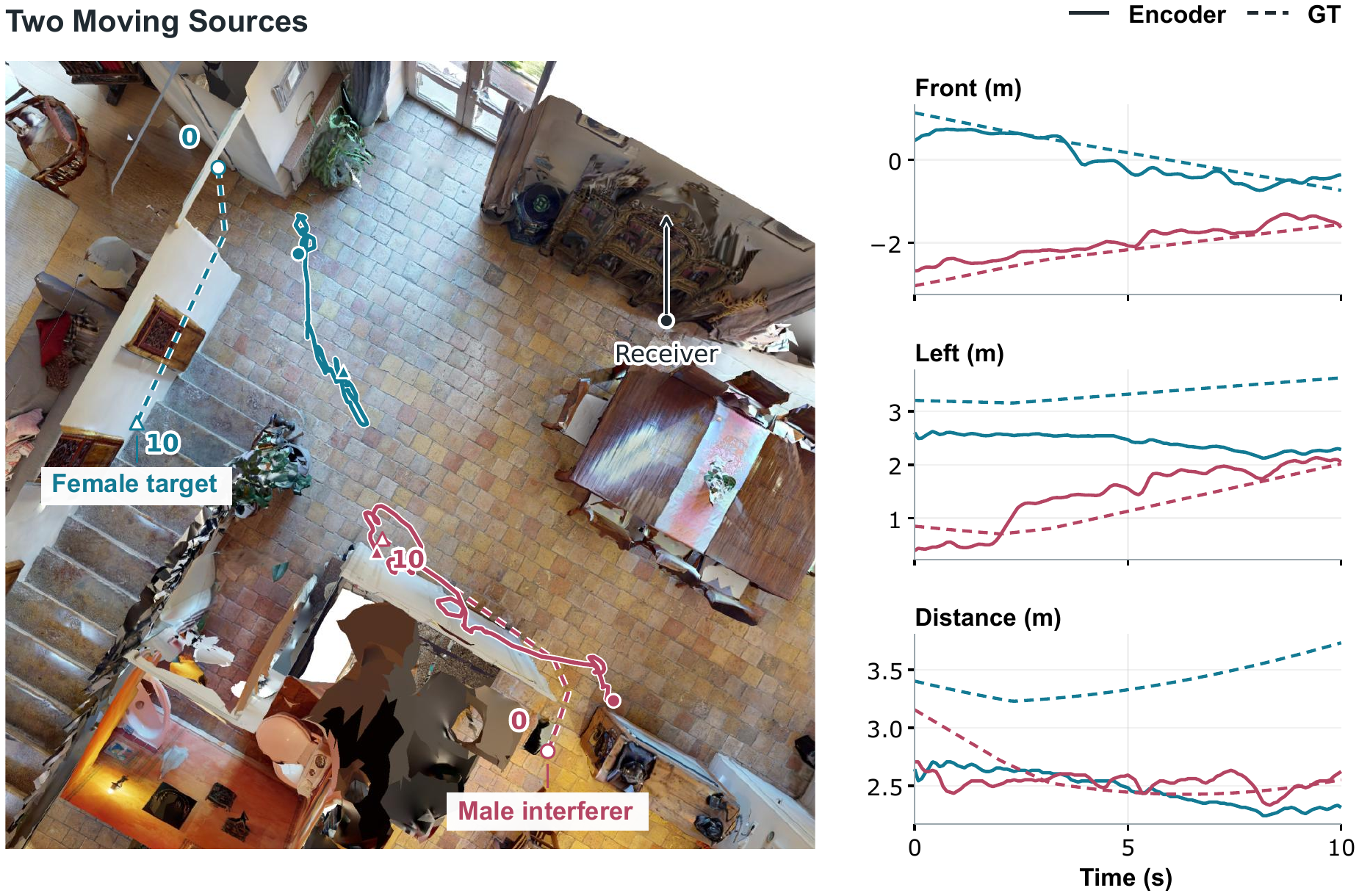}
  \caption{\textbf{Decoded states of two moving sources.}
  Colors distinguish the sources. Solid lines show decoded
  trajectories and dashed lines show ground truth.}
  \label{fig:case_state_dual}
\end{figure*}

\begin{figure*}[!tp]
  \centering
  \includegraphics[width=\linewidth,height=0.4\textheight,keepaspectratio]{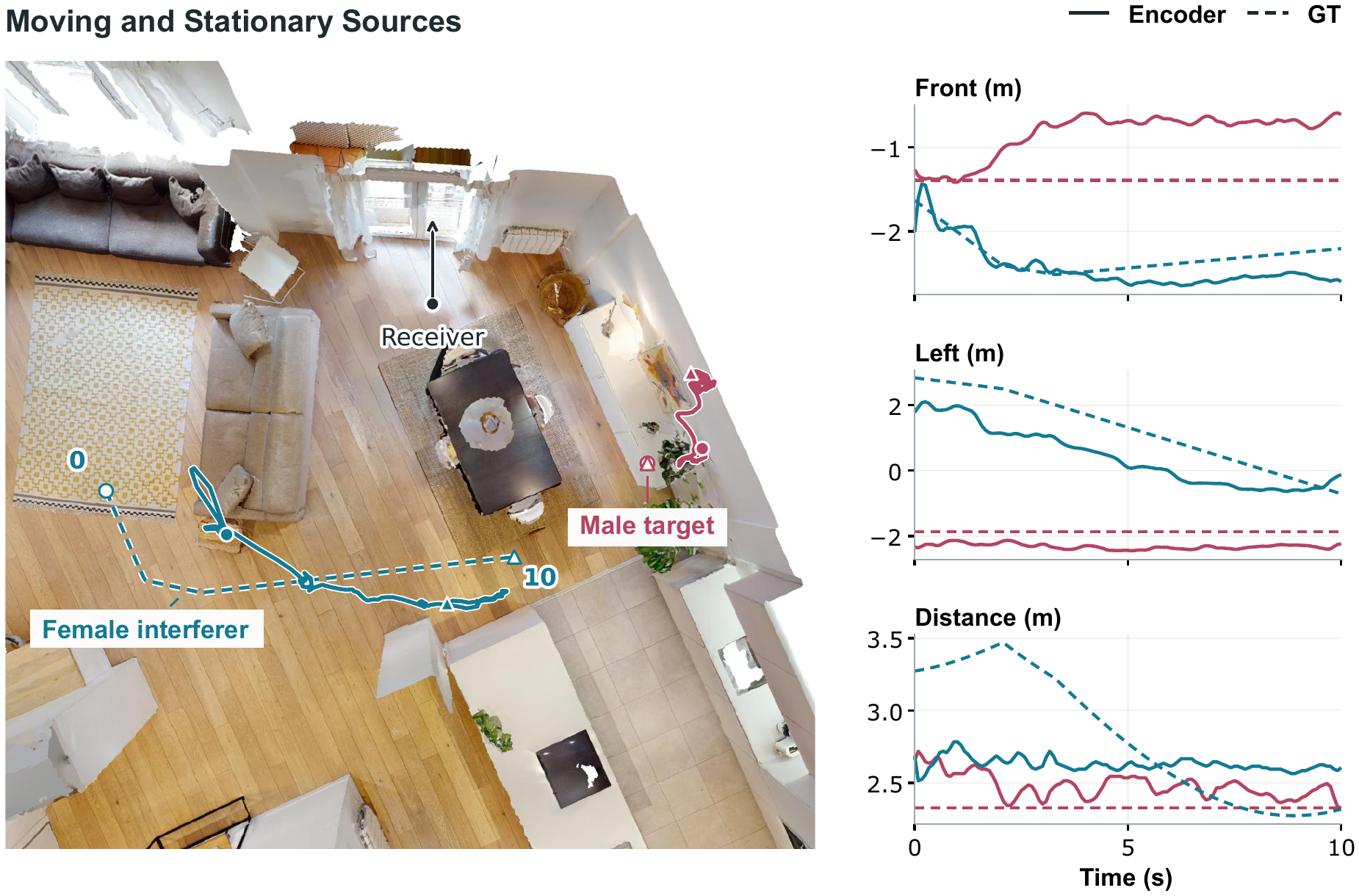}
  \caption{\textbf{Decoded states of a moving source
  and a stationary source.}
  Colors distinguish the sources. Solid lines show decoded
  trajectories and dashed lines show ground truth.}
  \label{fig:case_state_mixed}
\end{figure*}

\paragraph{Separating two moving sources.}
Figure~\ref{fig:case_state_dual} shows that two ground-truth trajectories exhibit opposing trends
in their front coordinates: one decreases over time,
while the other increases.
The decoded trajectories preserve these respective trends
in separate slots.

\paragraph{Distinguishing moving and stationary sources.}
In Figure~\ref{fig:case_state_mixed}, the decoded left
coordinate changes substantially for the moving source,
whereas the stationary source's lateral estimate remains
nearly constant.
These patterns appear in separate source slots before language generation.
% These contrasting temporal patterns appear in separate
% source slots before language generation, distinguishing
% lateral movement from a stable lateral position within
% the same acoustic scene.

\subsection{Tracking Across Speech Pauses}
\label{app:case_pauses}

\begin{table}[!htbp]
\centering
\footnotesize
\setlength{\tabcolsep}{4pt}
\renewcommand{\arraystretch}{1.10}

\caption{\textbf{Source-state continuity across speech pauses.}
ADE (m) and ID retention (\%) are evaluated after speech resumes,
with pre-pause slot assignments fixed. ID retention uses distance
thresholds of 0.5 m and 0.25 m for stationary and moving interferers,
respectively.}
\label{tab:speech_pause}

\begin{tabularx}{\linewidth}{
@{}>{\raggedright\arraybackslash}Xcc@{}
}
\toprule
\textbf{Variant} & ADE$\downarrow$ & ID retention$\uparrow$\\
\midrule

\multicolumn{3}{@{}l}{\textit{Stationary interferer}}\\[2pt]
w/o transition MLP
& 0.82 & 0.00\\
w/o masked-state loss
& 0.66 & 9.09\\
\textbf{Full SonicWorld}
& \textbf{0.52} & \textbf{30.30}\\

\midrule

\multicolumn{3}{@{}l}{\textit{Moving interferer}}\\[2pt]
w/o transition MLP
& 0.33 & 14.81\\
w/o masked-state loss
& 0.34 & 25.93\\
\textbf{Full SonicWorld}
& \textbf{0.32} & \textbf{33.33}\\

\bottomrule
\end{tabularx}
\end{table}

\begin{figure*}[!htbp]
\centering
  \includegraphics[width=\linewidth]{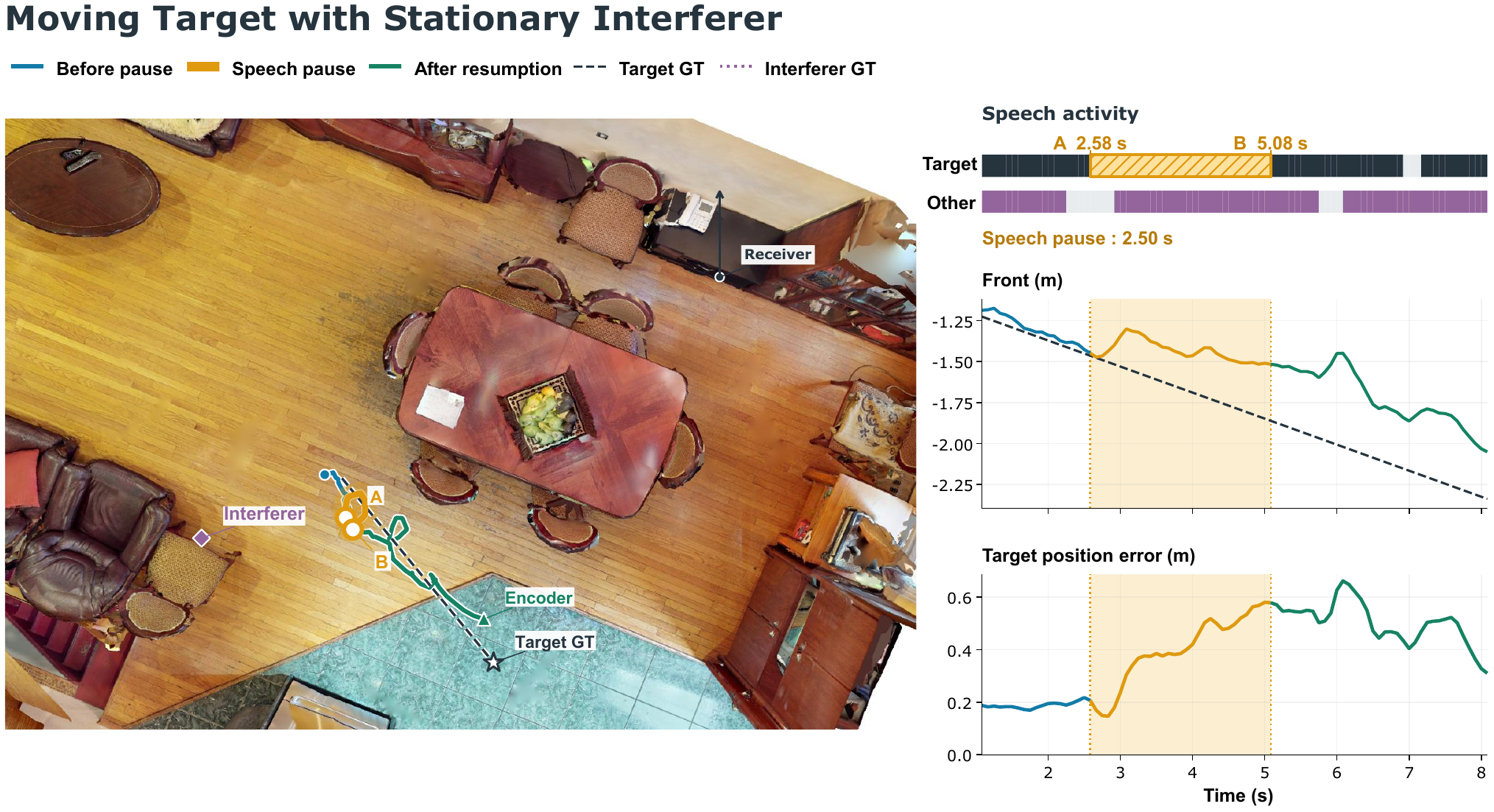}
\caption{\textbf{Tracking across a speech pause with a stationary interferer.}
Blue/gold/green show the decoded target trajectory before/during/after
the pause; A/B mark pause onset/resumption.
Dark dashed/purple dotted lines show target/interferer ground truth.
Right panels show speech activity, receiver-centered front coordinate,
and target position error.}
\label{fig:case_pause_stationary}
\end{figure*}

\begin{figure*}[!htbp]
  \centering

  \includegraphics[width=\linewidth]{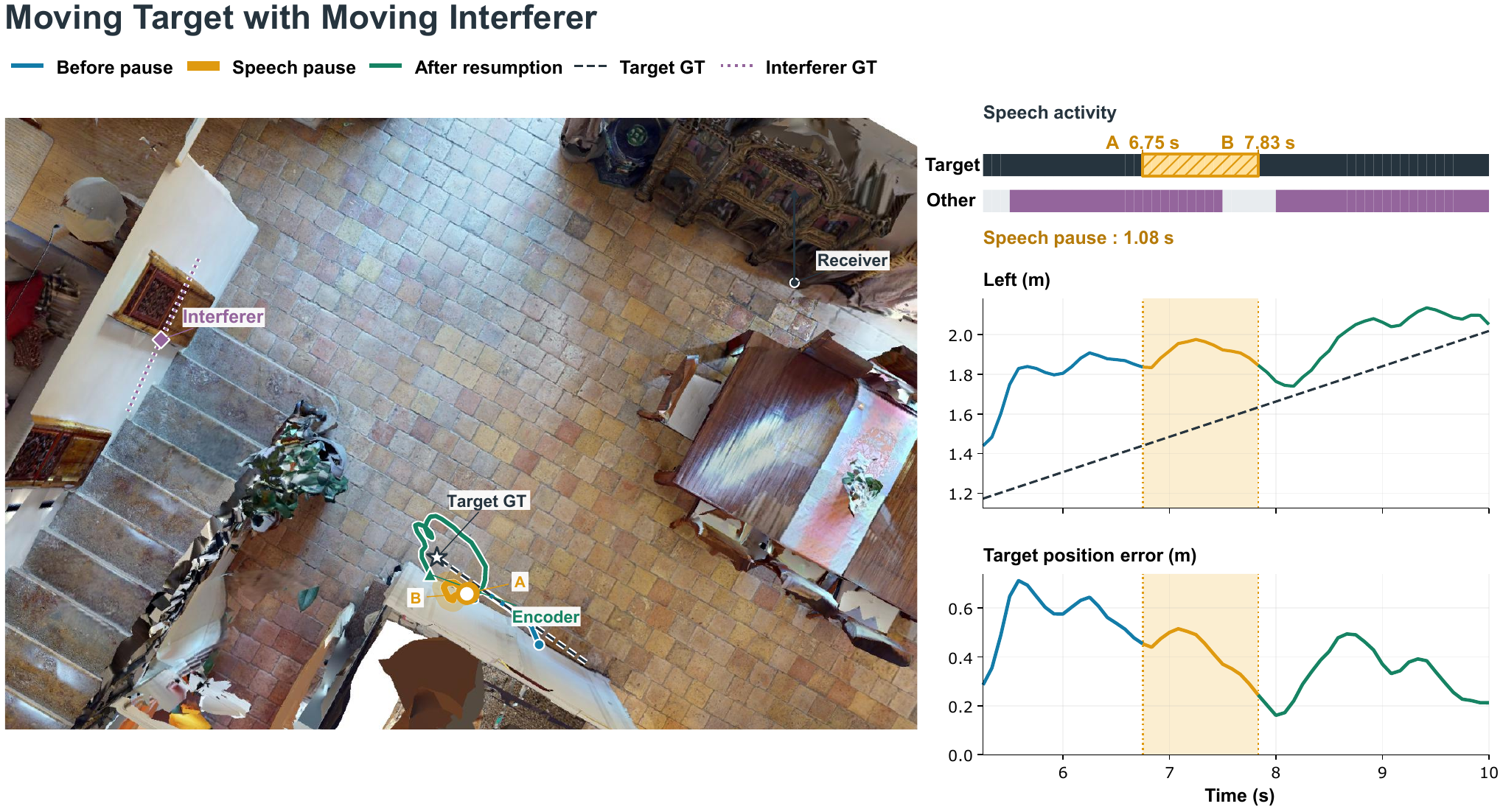}
  \caption{\textbf{Tracking across a speech pause
  with a moving interferer.}
  Both speakers continue moving during the target's 1.08\,s
  speech pause.
  Colors and markers follow
  Figure~\ref{fig:case_pause_stationary}.
  The coordinate panel shows the receiver-centered left position.}
  \label{fig:case_pause_moving}
\end{figure*}

We examine whether a source slot remains associated with
the same speaker after a speech pause.
Both examples contain a moving target that temporarily
stops speaking while continuing along its path, together with
a stationary or moving interferer.
Trajectories are decoded from the complete clip using the
source-state model with bidirectional refinement.

For each case, we assign predicted source slots to ground-truth
sources using their speech-active frames in the 1.5\,s preceding
the pause and keep this assignment fixed after resumption.
Let $s_0$ denote the assigned target slot and $\mathcal{T}$ the
target speech-active frames within 3\,s after resumption,
clipped at the end of the recording.
Given the predicted target position
$\hat{\mathbf p}_{t,s_0}$ and ground-truth position
$\mathbf p_t$, we compute average displacement error (ADE):
\[
\mathrm{ADE}=
\frac{1}{|\mathcal{T}|}
\sum_{t\in\mathcal{T}}
\|\hat{\mathbf p}_{t,s_0}-\mathbf p_t\|_2 .
\]
We additionally report ID retention, where $r_t=1$ if the
target remains uniquely matched to the assigned slot at frame
$t$, and $r_t=0$ otherwise:
\[
\mathrm{ID\ retention}(\%)=
\frac{100}{|\mathcal{T}|}
\sum_{t\in\mathcal{T}}r_t .
\]
Frame-level matching uses a distance threshold of 0.5\,m and
0.25\,m for stationary and moving interferers, respectively.

Table~\ref{tab:speech_pause} shows that the full model achieves
the lowest post-resumption ADE and highest ID retention in both
examples. Removing either component substantially worsens ADE
with a stationary interferer, while a moving interferer mainly
reveals differences in ID retention.
These case-level results support the role of the transition MLP
and masked-state loss in maintaining source-state continuity
across speech pauses.

\paragraph{Stationary interferer.}
In Figure~\ref{fig:case_pause_stationary}, the target pauses
from 2.58 to 5.08\,s while the interferer remains
stationary.
After speech resumes, the slot matched before the pause
follows the target's decreasing front coordinate.
The decoded trajectory retains a positional lag, with a
post-resumption ADE of 0.52\,m.
Thus, the same slot captures the target's continuing
front--back movement after the 2.50\,s speech interruption.

\paragraph{Moving interferer.}
In Figure~\ref{fig:case_pause_moving}, both speakers move,
and the target pauses from 6.75 to 7.83\,s.
Following this 1.08\,s interruption, the same slot captures
the target's overall increase in the left coordinate.
Its post-resumption ADE is 0.32\,m, evaluated on target-active
frames during the remaining 2.17\,s of the clip.

In both examples, the slot matched to the target before the
pause remains closer to the target than to the interferer
at every evaluated post-resumption frame.
These observations illustrate continuity of the target-associated
spatial readout across interrupted speech.

% \begin{figure}[!htbp]
%   \centering
%   \includegraphics[width=\linewidth]{figures/case_qa_000319.pdf}
%   \caption{\textbf{Movement predictions for a single moving source.}}
%   \label{fig:case_study_turns}
% \end{figure}

% \begin{figure}[!htbp]
%   \centering
%   \includegraphics[width=\linewidth]{figures/case_qa_000428.pdf}
%   \caption{\textbf{Movement predictions for a female target
%   with a moving interferer.}}
%   \label{fig:case_study_interference_female}
% \end{figure}

% Figures~\ref{fig:case_study_turns_interferer} show targets that
% turn during the queried interval. The movement label describes
% the net displacement between the interval endpoints.
% For the single moving source, SonicWorld correctly predicts
% Front-Left, recovering both components of the displacement.
% Spatial-Omni predicts Front, while JAEGER (Classical-IV)
% predicts Back-Right.
% With a stationary interferer, SonicWorld correctly predicts
% Back-Left; Spatial-Omni returns Stationary, and JAEGER
% (Classical-IV) returns Left.
% Across both cases, SonicWorld identifies the target's
% interval-specific displacement along a turning path.

% We examine source-specific motion reasoning at two levels:
% the final answers produced by SonicWorld and the trajectories
% decoded from its persistent source states. The cases progress
% from a single moving speaker to mixtures containing moving
% and stationary speakers.

\FloatBarrier
\section{Questions and Prompts}
\label{app:prompts}
This section presents the task instructions and complete question templates used for Sonic-Bench and the two external evaluations.

\subsection{Sonic-Bench}
\label{app:sonic_prompts}
The five question templates below identify the target voice and include the required answer format.
The placeholders \texttt{\{gender\}}, \texttt{\{time\}}, \texttt{\{start\}}, and \texttt{\{end\}} are instantiated for each question, with times written to two decimal places and the movement interval spanning one second.
SonicWorld places its input instruction after the audio and \texttt{<world>} spatial-token block and before the task-specific question.

\begin{evaluationprompt}{SonicWorld input instruction for Sonic-Bench}
You are an audio spatial reasoning assistant. Use the raw mono audio semantics and the complete source-centric dynamic acoustic world enclosed by <world> tags. The world contains persistent anonymous source slots across the full clip. Use the question's source identity, timestamp, and interval to select the relevant slot. Follow the requested answer format exactly and output no explanation.
\end{evaluationprompt}

\begin{evaluationprompt}{Sonic-Bench question: Position direction}
At {time} seconds, which direction is the {gender} voice coming from? Answer with exactly one label from: Front, Front-Left, Left, Back-Left, Back, Back-Right, Right, Front-Right.
\end{evaluationprompt}

\begin{evaluationprompt}{Sonic-Bench question: Azimuth}
At {time} seconds, what is the horizontal azimuth of the {gender} voice relative to the listener? Answer exactly as a number in degrees.
\end{evaluationprompt}

\begin{evaluationprompt}{Sonic-Bench question: Elevation}
At {time} seconds, what is the elevation of the {gender} voice relative to the listener? Answer exactly as a number in degrees.
\end{evaluationprompt}

\begin{evaluationprompt}{Sonic-Bench question: Distance}
At {time} seconds, how far away is the {gender} voice? Answer exactly in this format: distance: <meters> m.
\end{evaluationprompt}

\begin{evaluationprompt}{Sonic-Bench question: Movement direction}
Between {start} and {end} seconds, which direction did the {gender} sound source move relative to the listener's fixed orientation? Answer with exactly one label from: Front, Front-Left, Left, Back-Left, Back, Back-Right, Right, Front-Right, Stationary.
\end{evaluationprompt}

\subsection{Spatial LibriSpeech Questions}
SLS provides spatial annotations rather than question--answer pairs; we construct the following four questions for each retained single-source speech clip.

\begin{evaluationprompt}{SLS question: Position direction}
Which direction is the sound source coming from? Answer with exactly one label from: Front, Front-Left, Left, Back-Left, Back, Back-Right, Right, Front-Right.
\end{evaluationprompt}

\begin{evaluationprompt}{SLS question: Azimuth}
What is the horizontal azimuth of the sound source relative to the listener? Answer exactly as a number in degrees.
\end{evaluationprompt}

\begin{evaluationprompt}{SLS question: Elevation}
What is the elevation of the sound source relative to the listener? Answer exactly as a number in degrees.
\end{evaluationprompt}

\begin{evaluationprompt}{SLS question: Distance}
How far away is the sound source from the listener? Answer exactly in this format: distance: <meters> m.
\end{evaluationprompt}

\subsection{SpatialSceneQA  Questions}
\begingroup\emergencystretch=2em\relax
Azimuth and elevation are requested separately for both source-count subsets.
In dual-source clips, \texttt{\{target\_description\}} is ``male voice'' or ``female voice'', as specified by the original question.
\par\endgroup

\begingroup
\renewcommand{\evaluationprompttextcolor}{black}
\begin{evaluationprompt}{SpatialSceneQA  question: Single-source azimuth}
What is the horizontal azimuth of the sound source relative to the listener? Answer exactly as a number in degrees.
\end{evaluationprompt}

\begin{evaluationprompt}{SpatialSceneQA  question: Single-source elevation}
What is the elevation of the sound source relative to the listener? Answer exactly as a number in degrees.
\end{evaluationprompt}

\begin{evaluationprompt}{SpatialSceneQA  question: Dual-source azimuth}
What is the horizontal azimuth of the {target_description} relative to the listener? Answer exactly as a number in degrees.
\end{evaluationprompt}
% \vspace{-8pt}
\begin{evaluationprompt}{SpatialSceneQA  question: Dual-source elevation}
What is the elevation of the {target_description} relative to the listener? Answer exactly as a number in degrees.
\end{evaluationprompt}
\endgroup

\end{document}